\documentclass[pre,preprint,showpacs,superscriptaddress]{revtex4}
\usepackage[dvips]{graphicx}
\usepackage{times,epsfig,amssymb}

\begin{document}
\def\be{\begin{equation}}
\def\ee{\end{equation}}
\def\bea{\begin{eqnarray}}
\def\eea{\end{eqnarray}}

\title{Spin Glass Phase Transition on Scale-Free Networks}
\author{D.-H. Kim}
\affiliation{School of Physics and Center for Theoretical Physics,
Seoul National University, Seoul 151-747, Korea}
\author{G. J. Rodgers}
\affiliation{Department of Mathematical Sciences, Brunel
University, Uxbridge, Middlesex, UB8 3PH, United Kingdom}
\author{B. Kahng}
\author{D. Kim}
\affiliation{School of Physics and Center for Theoretical Physics,
Seoul National University, Seoul 151-747, Korea}
\date{\today}

\begin{abstract}
We study the Ising spin glass model on scale-free networks
generated by the static model using the replica method. Based on
the replica-symmetric solution, we derive the phase diagram
consisting of the paramagnetic (P), ferromagnetic (F), and spin
glass (SG) phases as well as the Almeida-Thouless line as
functions of the degree exponent $\lambda$, the mean degree $K$,
and the fraction of ferromagnetic interactions $r$.
To reflect the inhomogeneity of vertices, we modify the
magnetization $m$ and the spin glass order parameter $q$ with
vertex-weights. The transition temperature $T_c$ ($T_g$)
between the P-F (P-SG) phases and the critical behaviors
of the order parameters are found analytically. When
$2 < \lambda < 3$, $T_c$ and $T_g$ are infinite, and the system is
in the F phase or the mixed phase for $r > 1/2$, while it is
in the SG phase at $r=1/2$. $m$ and $q$ decay as power-laws with
increasing temperature with different $\lambda$-dependent exponents.
When $\lambda > 3$, the $T_c$ and $T_g$ are finite and
related to the percolation threshold. The critical exponents
associated with $m$ and $q$ depend on $\lambda$ for
$3 < \lambda < 5$ ($3 < \lambda < 4$) at the P-F (P-SG)
boundary.
\end{abstract}

\maketitle

\section{Introduction}
Recently, considerable effort has been devoted to understanding
complex systems by means of networks~\cite{Strogatz01,Albert02,
Dorogovtsev02a,Newman03}. An emerging phenomenon in real-world
complex networks is a scale-free (SF) behavior in the degree
distribution, $P_{d}(k) \sim k^{-\lambda}$, where the degree $k$ is
the number of edges connected to a given vertex and $\lambda$ is
the degree exponent~\cite{Barabasi99}. Due to the heterogeneity of
degree, many physical problems on SF networks exhibit distinct
features from those in Euclidean space. For example, the critical
behavior of the ferromagnetic Ising model on SF networks exhibits
an anomalous behavior depending on the degree exponent
$\lambda$~\cite{Aleksiejuk01,Dorogovtsev02b,Leone02,Bianconi02,herrero}.
While the critical behaviors are of the mean field type for
$\lambda > 5$, they exhibit an anomalous scaling for $3 < \lambda
<5$. Moreover, the magnetization, ${\bar m}$, decreases with
increasing temperature as ${\bar m} \sim T^{-1/(3-\lambda)}$
for $2 < \lambda < 3$, and so on
\cite{Dorogovtsev02b,Leone02}. The Ising spin system on the
complex networks, besides being of theoretical interest, can be
used to describe various real world phenomena. For example, the
two Ising spin states may represent two different opinions in a
society. Depending on the interaction strength between neighbors,
the overall system can be in a single or mixed opinion states,
corresponding to the ferromagnetic or paramagnetic phase, respectively.

In complex systems, such a description with only ferromagnetic
interactions may not be sufficient in certain circumstances. In
social systems, for example, the relationship between two
individuals can be friendly or unfriendly. In biological systems,
two genes can respond to an external perturbation coherently or
incoherently in microarray assay. For such cases, the spin glass
model is then more relevant to account for such competing
interactions. Recently, the spin glass problem has been studied on
the small world network proposed by Watts and Strogatz
\cite{Watts98} through both the replica method and the cavity
method \cite{Wemmenhove04}. Since SF networks are ubiquitous in
nature, here we study the spin glass model on SF networks.

The spin glass problem in the Euclidean space has been studied for
a long time by various methods
\cite{Binder86,Mezard87a,Fischer91,Mydosh93}. Most of the studies
for spin glasses have concentrated on regular lattices or the
infinite-range interaction model on fully-connected graphs, for
example, the Sherrington-Kirkpatrick (SK) model
\cite{Sherrington75}. To achieve our goal here, we follow the
study of the {\it dilute} Ising spin glass model with
infinite-range interactions, first performed by Viana and Bray (VB)
\cite{Viana85,Kanter87,Mezard87b,Mottishaw87,Wong88,Monasson98},
because the model is equivalent to the Ising spin glass problem on
the random graph proposed by Erd\H{o}s and R\'enyi
(ER)~\cite{Erdos59,Bollobas85}. The ER random graph may be
constructed as follows. The number of vertices $N$ is fixed and
assumed to be sufficiently large. Each vertex $i$ $(i=1,2,\ldots,
N)$ is assigned a weight $p_i$, which is given as $p_i=1/N$,
independent of the index $i$ for the ER model. Two vertices
$i$ and $j$ are selected with probabilities $p_i$ and $p_j$,
respectively, and if $i\ne j$, they are connected with an edge
unless the pair is already connected, which we call the fermionic
constraint.
This process is repeated $NK/2$ times. In such networks, the
probability that a given pair of vertices $(i,j)$ ($i\neq j$) is
not connected by an edge, denoted by $1-f_{ij}$, is given by
$(1-2p_ip_j)^{NK/2}\simeq \exp (-NK p_ip_j)$, while the connection
probability is
\be
f_{ij}=1-\exp (-N K p_ip_j).
\label{f_ij}
\ee
Since $p_ip_j=1/N^2$ for the ER graph, the fraction of bonds
present becomes $f_{ij}\approx K/N$ and the average number of
connected edges is $NK/2$. So $K$ is the mean degree, and
corresponds to $p$ of Ref.~\cite{Viana85}.

The SF network can be constructed through a generalization
of the above to the case where the vertex-weights are given by
\begin{equation}
p_{i}=\frac{i^{-\mu}}{\zeta_{N}(\mu)}
\label{p_i},
\end{equation}
where $\mu$ is a control parameter in the range $[0,1)$, and
$\zeta_{N}(\mu) \equiv \sum_{j=1}^{N} j^{-\mu}\approx
{N^{1-\mu}}/(1-\mu)$. Then the resulting network is a SF network
with a power-law degree distribution, $P_{d}(k) \sim
k^{-\lambda}$, with $\lambda = 1 + 1/\mu$. The model is called the
static model, where the name `static' originates from the fact
that the number of vertices is fixed from the beginning
\cite{Goh01}. This model has the advantage that many of its
theoretical quantities can be calculated
analytically~\cite{Lee04}. Note that since $NKp_ip_j \sim
N^{2\mu-1}/(ij)^{\mu}$ for finite $K$, when $0 < \mu < 1/2$
($\lambda > 3$),
\begin{equation}
f_{ij}\approx NKp_ip_j,
\label{f_ij_a}
\end{equation}
however, when $1/2 < \mu < 1$ ($2 < \lambda < 3$), $f_{ij}$
does not necessarily take the form of Eq.(\ref{f_ij_a}),
but it is given as
\begin{eqnarray}
f_{ij}\approx\left \{
\begin{array}{lllll}
1 & & {\rm when} & & ij \ll N^{2-1/\mu}, \\
NKp_ip_j & & {\rm when} & & ij \gg N^{2-1/\mu}.
\end{array}
\right.
\label{f_ij_f}
\end{eqnarray}
This is due to the fermionic constraint that at most one edge
can be attached to a given pair of vertices. The mean degree
of a vertex $i$ is $NKp_i$ and the mean degree of the network
is $K$~\cite{Lee04}.

In this work, we study the Ising spin glass model defined
on the static model. In Sec.II, we introduce the
Hamiltonian of the spin glass system on the static model and
derive the free energy by using the replica method. We also
introduce physical quantities such as the magnetization, and the
spin glass order parameters in a modified form. In Sec.III, we
present the replica-symmetric solutions by using the SK-type
approximation, from which the phase diagram including the Almeida-Thouless
line and the critical behavior of the spin glass order parameters are
derived. In Sec.IV, we use the perturbative
approach to derive the phase diagram and the critical behaviors of the
order parameters, and compare them with those obtained from the
SK method. The final section is devoted to the conclusions and
discussion.

\section{The spin glass model}

We consider the Ising-type Hamiltonian,
\begin{equation}
\mathcal{H}=-\sum_{(i,j)\in G} J_{ij} s_{i} s_{j}~~(s_{i}=\pm 1),
\end{equation}
defined on a graph $G$ realized by the static model. $J_{ij}$ is
nonzero only when the vertices $i$ and $j$ are connected in $G$.
The network ensemble average for a given physical quantity $A$ is
taken as
\begin{equation}
\langle A \rangle_{K} = \sum_{G} P_{K}(G) A(G),
\end{equation}
where $P_{K}(G)$ is the probability of $G$ in the ensemble
and $\langle \cdots \rangle_K$ the average over different
graph configurations.
For the static model we consider here, it is given that
\begin{equation}
P_K(G)=\prod_{(i,j) \in G} f_{ij}\prod_{(i,j)\notin G}(1-f_{ij})
\end{equation}
with $f_{ij}=1-\exp (-N K p_{i}p_{j})$, $p_{i}$ being given in
Eq.(\ref{p_i}).

In the spin glass problem, the coupling strengths $\{J_{ij}\}$ are
also quenched random variables. We assume in this paper that
each $J_{ij}$ is given as $+J$ or $-J$ with probability $r$
and $1-r$, respectively, so that the coupling strength distribution
is given as
\begin{equation}
P_{r}(\{J_{ij}\})=\prod_{(i,j)\in G} \Big[ r\delta
(J_{ij}-J)+(1-r)\delta(J_{ij}+J) \Big].
\end{equation}
The case of $r=1$ ($r=1/2$) is pure ferromagnetic (fully
frustrated) one, and we consider $r$ in the range of $1/2 \le r
\le 1$ throughout this work. The average of a quantity $A$ with
respect to $P_{r}(\{J_{ij}\})$ is denoted as $\langle A
\rangle_{r}$. Thus the free energy is evaluated as $-\beta F =
\langle \langle \ln Z \rangle_{r} \rangle_{K}$ with $Z$ being the
partition function for a given distribution of $\{J_{ij}\}$ on a
particular graph $G$.

In this paper, the replica method is used to evaluate the free
energy, i.e., $-\beta F = \lim_{n \to 0} [\langle \langle Z^{n}
\rangle_{r} \rangle_{K}-1]/n$. To proceed, we evaluate the $n$-th
power of the partition function,
\begin{eqnarray}
\langle \langle Z^{n} \rangle_{r} \rangle_{K} &=&
\textrm{Tr}_{\{s^{\alpha}\}} \Big\langle \Big\langle \exp (\beta
\sum_{ (i,j)\in G} J_{ij} \sum_{\alpha=1}^{n}
s_{i}^{\alpha} s_{j}^{\alpha})
\Big\rangle_{r} \Big\rangle_{K} \nonumber \\
&=& \textrm{Tr}_{\{s^{\alpha}\}} \prod_{i<j} \Big\{ (1 - f_{ij}) +
f_{ij} \Big \langle \exp ( \beta J_{ij} \sum_{\alpha=1}^{n}
s_{i}^{\alpha} s_{j}^{\alpha} ) \Big \rangle_{r} \Big\} \nonumber
\\ &=& \textrm{Tr}_{\{s^{\alpha}\}} \exp \Big[ \sum_{i < j}\ln
\Big\{ 1 + f_{ij} \Big( \Big\langle \exp (\beta J_{ij}
\sum_{\alpha=1}^{n} s_{i}^{\alpha} s_{j}^{\alpha})-1
\Big\rangle_{r} \Big) \Big\} \Big], \label{z_n}
\end{eqnarray}
where the trace $\textrm{Tr}_{\{s^{\alpha}\}}$ is taken over all
replicated spins $s_{i}^{\alpha}=\pm 1$, $\alpha=1,\ldots,n$ is
the replica index, and $\beta=1/T$. We mention that the disorder
averages over $P_K(G)$ and $P_{r}(\{J_{ij}\})$ can be done
simultaneously since both types of disorders are independently
assigned to each edge of the fully-connected graph of order $N$.
The part inside the exponential in Eq.(\ref{z_n}) can be written
in the form,
\begin{eqnarray}
\sum_{i < j} \ln \Big\{ 1 + f_{ij} \Big( \Big\langle \exp (\beta
J_{ij} \sum_{\alpha=1}^{n} s_{i}^{\alpha} s_{j}^{\alpha})-1
\Big\rangle_{r} \Big) \Big\}&=& \sum_{i < j} NKp_{i}p_{j}
\Big\langle \exp (\beta J_{ij}\sum_{\alpha=1}^{n} s_{i}^{\alpha}
s_{j}^{\alpha}) -1 \Big\rangle_{r}+R \nonumber \\
\label{argu}
\end{eqnarray}
where $R$ stands for the remainder which are of higher order in
$K$. It is shown in APPENDIX A that for finite $K$,
Eq.(\ref{argu}) is $\mathcal{O}(N)$ while $R$ is at most
$\mathcal{O}(1)$ for $\lambda >3$ and
$\mathcal{O}(N^{3-\lambda}\ln N)$ for $2 < \lambda <3$, so
that it can be neglected in the free energy calculation.

Once $R$ in Eq.(\ref{argu}) can be neglected, we can proceed as
in VB~\cite{Viana85}. By using the relation,
\begin{eqnarray}
\Big\langle \exp(\beta J_{ij}\sum_{\alpha=1}^{n} s_{i}^{\alpha}
s_{j}^{\alpha}) \Big\rangle_{r} = \Big\langle \prod_{\alpha} \Big[
\cosh(\beta J_{ij})(1 + s_{i}^{\alpha} s_{j}^{\alpha} \tanh(\beta
J_{ij})) \Big] \Big\rangle_{r},
\end{eqnarray}
in Eq.(\ref{argu}) and applying the Hubbard-Stratonovich identity,
Eq.(\ref{z_n}) is reduced to the form
\begin{equation}
\langle \langle Z^n \rangle_{r}
\rangle_{K} = \int d\mathbf{q} \exp \{-N n\beta f(\mathbf{q})\}.
\label{integral}
\end{equation}
The intensive free energy $f \{\mathbf{q} \} (\equiv F/Nn)$ in the
thermodynamic limit ($N \to \infty$) then becomes
\begin{eqnarray}
n \beta f \{\mathbf{q} \} = \frac{K \mathbf{T}_{1}}{2}
\sum_{\alpha} q_{\alpha}^{2} + \frac{K \mathbf{T}_{2}}{2}
\sum_{\alpha < \beta} q_{\alpha \beta}^{2} + \frac{K
\mathbf{T}_{3}}{2} \sum_{\alpha < \beta < \gamma} q_{\alpha \beta
\gamma}^{2} + \cdots -\frac{1}{N} \sum_{i} \ln
\textrm{Tr}_{\{s_i^{\alpha}\}} \exp X_{i}, \nonumber \\
\label{free_energy}
\end{eqnarray}
where
\begin{eqnarray}
X_{i}= N K \mathbf{T}_{1} p_i \sum_{\alpha}
q_{\alpha}s_{i}^{\alpha}+ N K \mathbf{T}_{2} p_i \sum_{\alpha <
\beta} q_{\alpha \beta} s_{i}^{\alpha} s_{i}^{\beta} + N K
\mathbf{T}_{3} p_i \sum_{\alpha < \beta < \gamma} q_{\alpha \beta
\gamma} s_{i}^{\alpha} s_{i}^{\beta} s_{i}^{\gamma}+\cdots,
\label{x_i}
\end{eqnarray}
and
\begin{eqnarray}
\mathbf{T}_{l}(T) \equiv \langle \cosh^n \beta J_{ij} \tanh^{l}
\beta J_{ij} \rangle_r  \stackrel{n \to 0}{\longrightarrow}
[r+(-1)^{l}(1-r)]\tanh^{l}\beta J ~~~~(l=1,2,\ldots).
\label{t_l}
\end{eqnarray}
$\textrm{Tr}_{\{s_i^{\alpha}\}}$ is the trace over the replicated
spins at vertex $i$ and the $N \to \infty$ limit is to be
implicitly understood to the expression $\frac{1}{N} \sum_{i}$.
The elements of a set $\{\mathbf{q} \}$, $q_{\alpha}$, $q_{\alpha
\beta}$, $q_{\alpha \beta \gamma}$, etc., defined as
\begin{equation}
q_{\alpha}=\sum_{i} p_{i} \langle s_{i}^{\alpha}\rangle_{i}, ~~~~
q_{\alpha \beta}=\sum_{i} p_{i} \langle s_{i}^{\alpha}
s_{i}^{\beta} \rangle_{i}, ~~~~ q_{\alpha \beta \gamma}=\sum_i p_i
\langle s_i^{\alpha} s_i^{\beta} s_i^{\gamma}\rangle_i,
~~~~\hbox{etc.} \label{op}
\end{equation}
are the order parameters of the spin glass system, called the
magnetization, the spin glass order parameter, and so on.
The average is evaluated through $\langle A \rangle_{i} \equiv
\textrm{Tr}_{\{s_i^{\alpha}\}} A \exp
X_{i}/\textrm{Tr}_{\{s_i^{\alpha}\}} \exp X_{i}$.
Note that unlike the case of the ER random graph,
the order parameters are summed with weight $\{p_i\}$ in
Eq.(\ref{op}) due to the inhomogeneity of the SF networks. For
the ER case however, $p_{i}=1/N$ and it becomes that ${\bar
q_{\alpha}}=\sum_i \langle s_i^{\alpha} \rangle_i/N$,
${\bar q_{\alpha \beta}}=\sum_i\langle s_{i}^{\alpha}s_{i}^{\beta}
\rangle_{i}/N$, ${\bar q_{\alpha \beta \gamma}}=\sum_i \langle
s_i^{\alpha}s_i^{\beta} s_i^{\gamma}\rangle_i/N$, and so on
\cite{Viana85}. To distinguish, we use bar notation for the
unweighted cases.

Here we consider the replica symmetry (RS) in which spins with
different replica index are indistinguishable, and we invoke two
methods to determine the phase boundaries of the ferromagnetic
(F), paramagnetic (P) and spin glass (SG) phases and the
temperature dependences of the order parameters. The first is the
approach similar in spirit to SK in which higher-order terms than
$q_{\alpha \beta}$ in Eqs.(\ref{free_energy}) and (\ref{x_i}) are
neglected. In this method, the remaining two order parameters as
well as the Almeida-Thouless line can be obtained for all
temperatures. The second is the perturbative approach used in VB.
In this case, we expand the term of $\ln {\rm Tr}
\exp X_i$ in Eq.(\ref{free_energy}) up to appropriate orders,
and the order parameters $q_{\alpha}$, $q_{\alpha \beta}$,
$q_{\alpha \beta \gamma}$ and $q_{\alpha \beta \gamma
\delta}$ are explicitly calculated. Through the perturbative
approach, we can find that the contributions by higher order terms
such as $q_{\alpha \beta \gamma}$ are negligible compared with
those by $q_{\alpha}$ and $q_{\alpha \beta}$ near the phase
transition points. Thus, the two methods produce identical
results for the phase boundaries and the
same critical behaviors near the transition points for the two order
parameters, $q_{\alpha}$ and $q_{\alpha \beta}$.

\section{The Sherrington-Kirkpatrick approach}

\subsection{The replica symmetric free energy}

We first study the RS solution~\cite{Sherrington75} and obtain the
phase boundaries of P, F and SG. For simplicity, the RS
magnetization and the RS spin glass order parameter are denoted as
$m(=q_{\alpha})$ and $q(=q_{\alpha \beta})$, respectively, and the
free energy expression Eq.(\ref{free_energy}) is truncated at the
order of $q$. Then the RS free energy is rewritten as
\begin{equation}
n \beta f(m,q) = \frac{K \mathbf{T}_{1}}{2} n m^{2} + \frac{K
\mathbf{T}_{2}}{2}\frac{n(n-1)}{2} q^{2} -\frac{1}{N} \sum_{i} \ln
\mathcal{Z}_{i}
\end{equation}
with
\begin{equation}
\mathcal{Z}_{i} = \textrm{Tr}_{\{s_i^{\alpha}\}} \exp \Big\{N K
\mathbf{T}_{1}p_i m \sum_{\alpha}s_{i}^{\alpha}+ N K
\mathbf{T}_{2}p_i q \frac{(\sum_{\alpha}
s_{i}^{\alpha})^{2}-n}{2}\Big\}.
\end{equation}
By using the Hubbard-Stratonovich identity, ${\cal Z}_i$ can
be rewritten as
\begin{equation}
{\cal Z}_i=\exp \{-\frac{n}{2} N K \mathbf{T}_{2}p_i q \} \int
\mathcal{D}z [2 \cosh \eta_{i}(z)]^{n},
\end{equation}
where $\int \mathcal{D}z \cdots \equiv \frac{1}{\sqrt{2\pi}}
\int_{-\infty}^{\infty} dz ~e^{-z^2/2} \cdots$ and $\eta_{i}(z)
\equiv NK \mathbf{T}_{1} p_{i} m + z \sqrt{NK \mathbf{T}_{2} p_{i}
q}$. Then in the limit of $n\rightarrow 0$, the RS free energy
becomes
\begin{equation}
\beta f(m,q) =\frac{1}{2} K \mathbf{T}_{1}
m^{2}+\frac{1}{2} K \mathbf{T}_{2}q -\frac{1}{4} K \mathbf{T}_{2}
q^{2}-\int \mathcal{D}z \frac{1}{N} \sum_{i=1}^{N} \ln [2\cosh
\eta_{i}(z)].
\label{free_energy_sk}
\end{equation}

By applying $(\partial f/\partial m)=0$ and $(\partial f/\partial
q)=0$ to the free energy, Eq.(\ref{free_energy_sk}), we obtain the
coupled self-consistent equations for $m$ and $q$ to be
\begin{equation}
m=\int \mathcal{D}z \sum_{i=1}^{N} p_{i} \tanh (NK \mathbf{T}_{1}
p_{i} m + z \sqrt{NK \mathbf{T}_{2} p_{i} q}), \label{self_m}
\end{equation}
and
\begin{equation}
q=\int \mathcal{D}z \sum_{i=1}^{N} p_{i} \tanh^{2} (NK
\mathbf{T}_{1} p_{i} m+z\sqrt{NK \mathbf{T}_{2}p_{i}q}).
\label{self_q}
\end{equation}
In Eqs.(\ref{self_m}) and (\ref{self_q}), we can see that $q$
cannot be zero unless both $m$ and $q$ are zero, while $m$ can be
zero even when $q \neq 0$ which defines the SG phase.

\subsection{The phase boundaries}

The P-F (P-SG) phase boundary is given as the temperature,
the Curie temperature $T_c$ (the spin glass phase transition
temperature $T_g$), where $m$ ($q$) starts to be nonzero.
We first consider the case of $\lambda > 3$. When $m$ and $q$ are
small, the free energy, Eq.(\ref{free_energy_sk}), is written as
\begin{eqnarray}
\beta f(m,q) &=& \frac{1}{2} K \mathbf{T}_{1} (1-NK\mathbf{T}_1
\sum_i p_i^2) m^{2} -\frac{1}{4} K \mathbf{T}_{2}
(1-NK\mathbf{T}_2 \sum_i p_i^2) q^{2} \nonumber \\
&+&\hbox{higher order terms.}
\end{eqnarray}

It is known that as $K$ increases, the static model undergoes
the percolation transition at
\be
K_p=\frac{1}{N\sum_i p_i^2}=
\frac{(\lambda-1)(\lambda-3)}{(\lambda-2)^2}.
\label{K_p}
\ee
Since $N\sum_i p_i^2 =(\langle k^2 \rangle_K-\langle k
\rangle_K)/ \langle k \rangle_K^2$
with $\langle k \rangle_K=K$ and $\langle k^2 \rangle_K$ denoting the
first and the second moments of the degree for a given mean degree
$K$, respectively, Eq.(\ref{K_p}) is equivalent to the condition
$\langle k^2 \rangle_K=2 \langle k \rangle_K$~\cite{Lee04,molloy,cohen}.
Thus one obtains that
\begin{eqnarray}
{\bf T}_{1}(T_c) &=& K_{p}/K ~~~~\hbox{for P-F,~~~~~~and} \label{t_c}\\
{\bf T}_{2}(T_g) &=& K_{p}/K ~~~~\hbox{for P-SG},\label{t_g}
\end{eqnarray}
where ${\bf T}_{1}(T)=(2r-1) \tanh (J/T)$ and ${\bf
T}_{2}(T)=\tanh^2 (J/T)$. Note that when $K/K_p < 1$, there is no
solution of Eqs.(\ref{t_c}) and (\ref{t_g}), implying that the
system is always in the P state. This is because the network has
an infinite component only for $K > K_p$.
When $r=1/2$, ${\bf T}_1=0$ and the phase diagram is rather simple.
The P-F transition does not occur, and the system is either in P
or SG phase whose boundary is given by Eq.(\ref{t_g}).
\begin{figure}
\includegraphics[angle=270, scale=0.25]{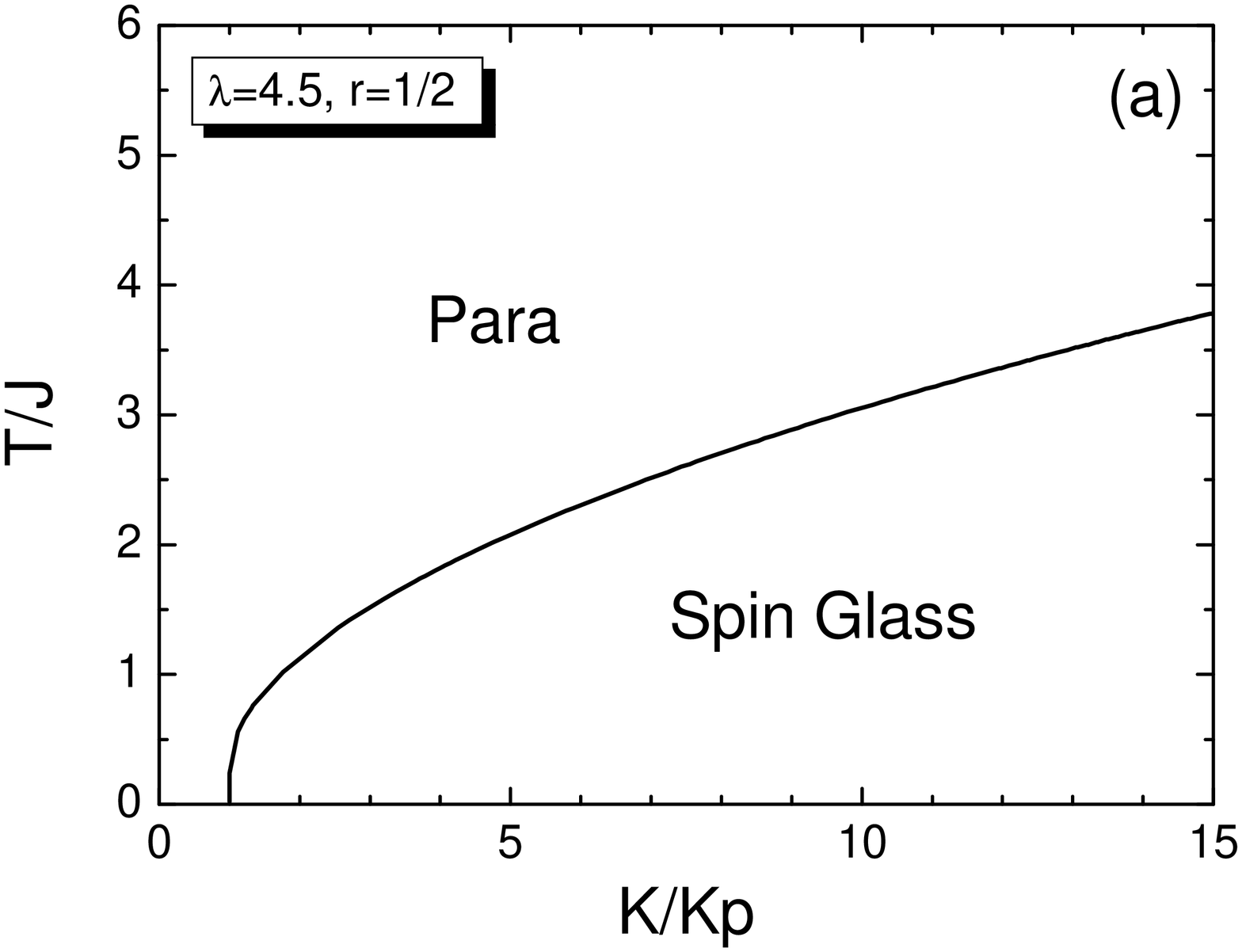}
\includegraphics[angle=270, scale=0.25]{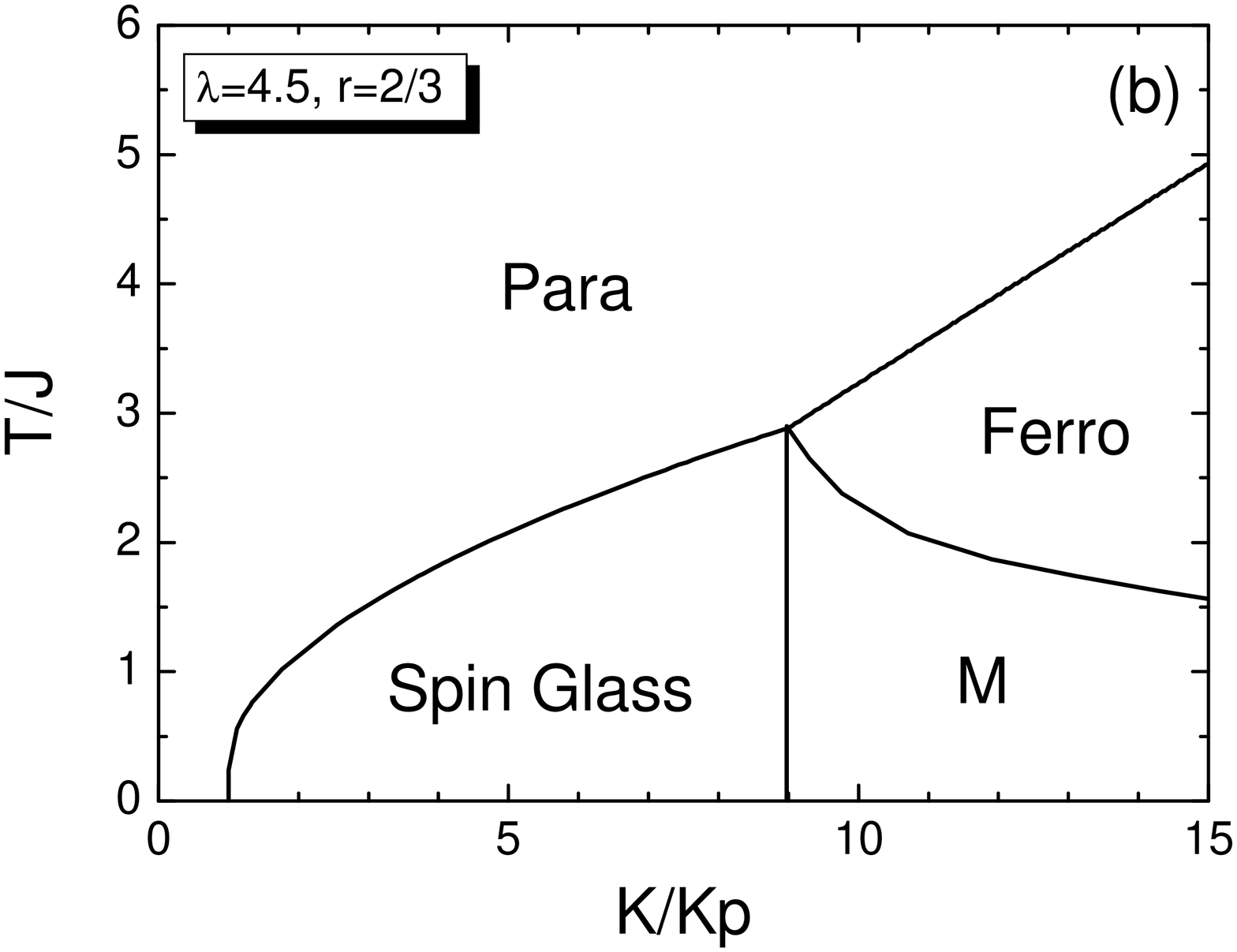}
\includegraphics[angle=270, scale=0.25]{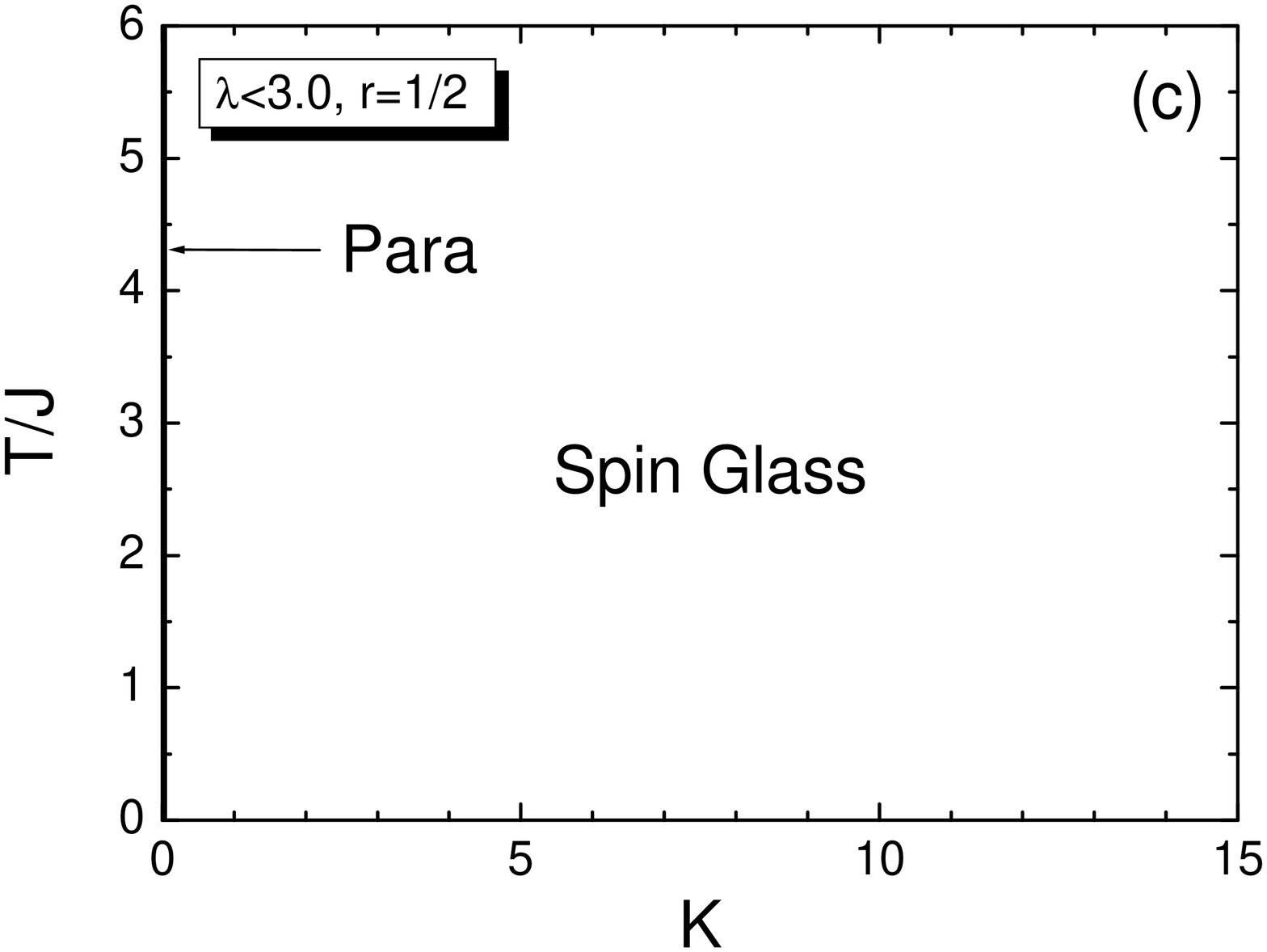}
\includegraphics[angle=270, scale=0.25]{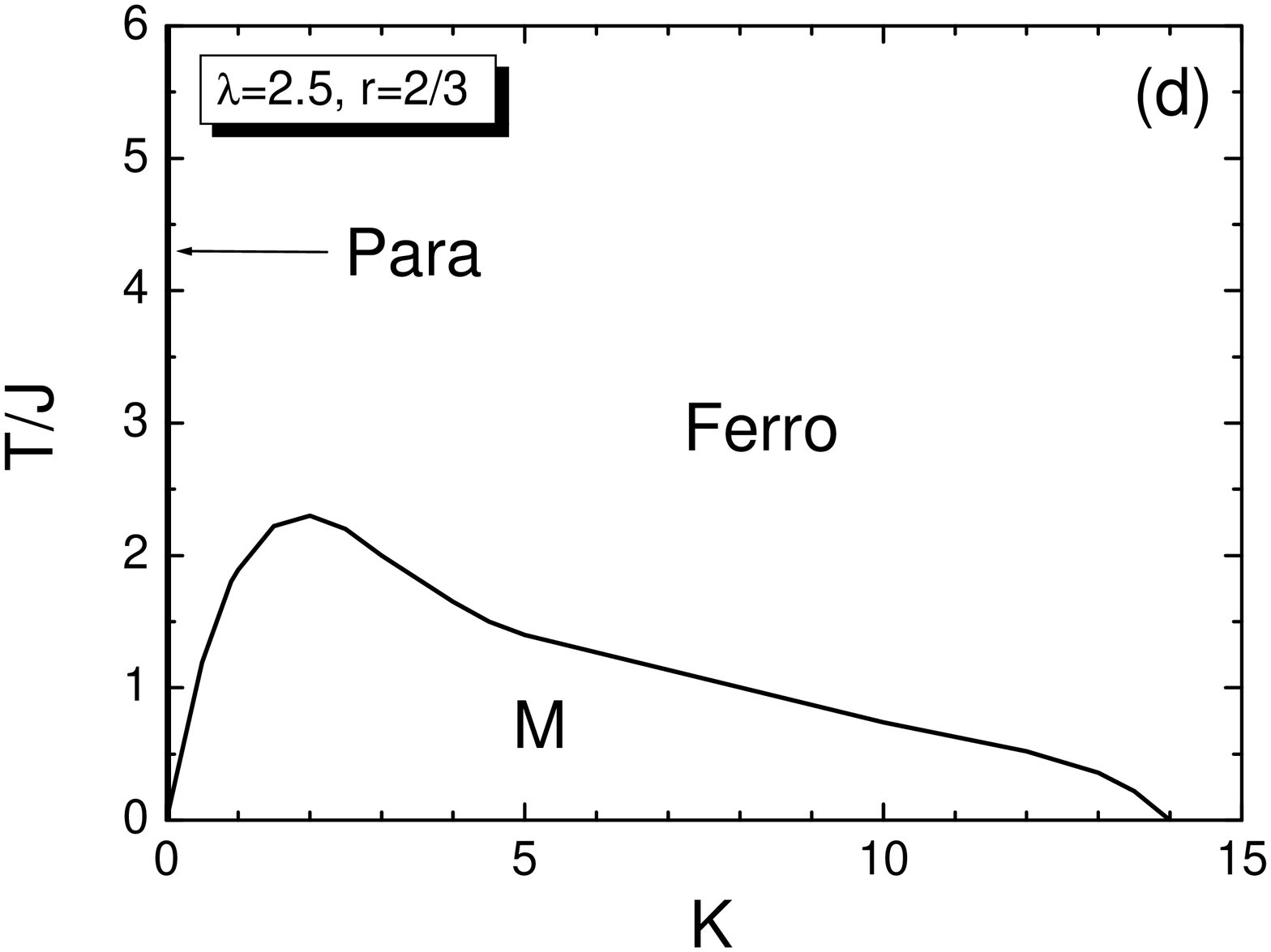}
\caption{The phase diagram in the $(K/K_{p},T/J)$ plane for
$\lambda=4.5(>3)$ with $r=1/2$ (a) and $\lambda=4.5(>3)$ with
$r=2/3$ (b), and the same in the $(K,T/J)$ plane for $\lambda<3$
with $r=1/2$ (c) and $\lambda=2.5(<3)$ with $r=2/3$ (d). Note that
$K_{p}=0$ for $2<\lambda<3$ in the thermodynamic limit.}
\label{fig:pd}
\end{figure}
FIG.{\ref{fig:pd}}(a) is the phase diagram in the $(K/K_{p},T/J)$
plane for the fully frustrated case ($r=1/2$) for $\lambda > 3$.
When $1/2 < r < 1$, both the F phase $(m \neq 0, q \neq 0)$ and
the SG phase $(m = 0, q \neq 0)$ appear.
FIG.\ref{fig:pd}(b) is the phase diagram for a partially
frustrated case with $r=2/3$ and $\lambda=4.5$, which is a
prototypical case of $1/2 < r < 1$ and $\lambda > 3$. For $K/K_{p}
< 1$, only the P phase appears, but for $K/K_p > 1$, several
phases exist.
There exists a multicritical point
$(K^{*}/K_p,T^{*}/J)$, where the P-SG-F phases merge, which is
determined to be
\begin{equation}
\Big(\frac{K^{*}}{K_p},\frac{T^{*}}{J}\Big)=
\Big(\frac{1}{(2r-1)^{2}},\frac{1}{\tanh^{-1}(2r-1)}\Big)
\label{t_multi}
\end{equation}
by setting $\mathbf{T}_{1}(T^{*})=\mathbf{T}_{2}(T^{*})=K_{p}/K^{*}$.
For $K_p < K < K^*$, the P phase goes into the SG phase, while
it goes into the F phase for $K > K^*$ as temperature is lowered.
As $r \to 1$, the multicritical point converges to $(1,0)$, indicating
that only the P-F phase transition occurs. As $r \to 1/2$, it
shifts to $(\infty, \infty)$, indicating that only the P-SG phase
transition occurs as shown in FIG.\ref{fig:pd}(a).

Besides the P, F, and SG phases, the mixed (M) phase is present,
which is defined as the re-entrant SG phase with nonzero
macroscopic ferromagnetic order, located below the F phase as
temperature is lowered~\cite{Fischer91,Mydosh93}. The SG-M phase
boundary is determined as the vertical straight line from the
multicritical point to $T/J=0$ \cite{Toulouse80}. The F-M phase
boundary is determined by the so-called Almeida-Thouless (AT)
line~\cite{Almeida78},
\begin{eqnarray}
(K \mathbf{T}_{2})^{-1}= \int \mathcal{D}z \sum_{i=1}^{N} N p_{i}^2~
\textrm{sech}^{4}(NK{\bf T}_1 p_i m +z\sqrt{NK{\bf T}_2 p_i q}),
\label{at}
\end{eqnarray}
which is obtained easily by multiplying vertex-weights to the AT
line formula of the SK model. $m$ and $q$ above
are the solutions of Eqs.(\ref{self_m}) and (\ref{self_q}). We
determine $T$ satisfying Eq.(\ref{at}) numerically. The F-M
boundary in FIG.\ref{fig:pd}(b) exhibits a fat-tail behavior,
implying that the M phase persists for large $K$. This AT line
is the phase boundary between the replica symmetric phase
and the replica-symmetry-broken one. Thus, Eq.(\ref{at}) indicates
the region where the replica-symmetric solution derived in the
following sections is valid. We also check the P-SG boundary from
Eq.(\ref{at}), which is the same as Eq.(\ref{t_g}).

Next we consider the case $2 < \lambda < 3$. In this range,
$K_{p} \sim N^{-(3-\lambda)/(\lambda-1)} \to 0$ as $N \to \infty$
and consequently $T_{c}$ and $T_{g} \to \infty$. Thus the whole
$(K, T/J)$ plane is covered with the ordered states.
FIG.{\ref{fig:pd}}(c) is the phase diagram for the fully
frustrated case ($r=1/2$) for $\lambda < 3$. The P phase appears
only for $K=0$, and the SG phase is located in the region $K>0$.
FIG.\ref{fig:pd}(d) deals with the case of $1/2< r < 1$ and
$\lambda < 3$. The P phase appears only at $K=0$, but for $K > 0$
the F and M phases appear and the F-M boundary is given by the AT line
(Eq.(\ref{at})). As $r \to 1$, the M phase disappears
and only the F phase appears in the region of $K > 0$.

\begin{figure}
\includegraphics[angle=270, scale=0.25]{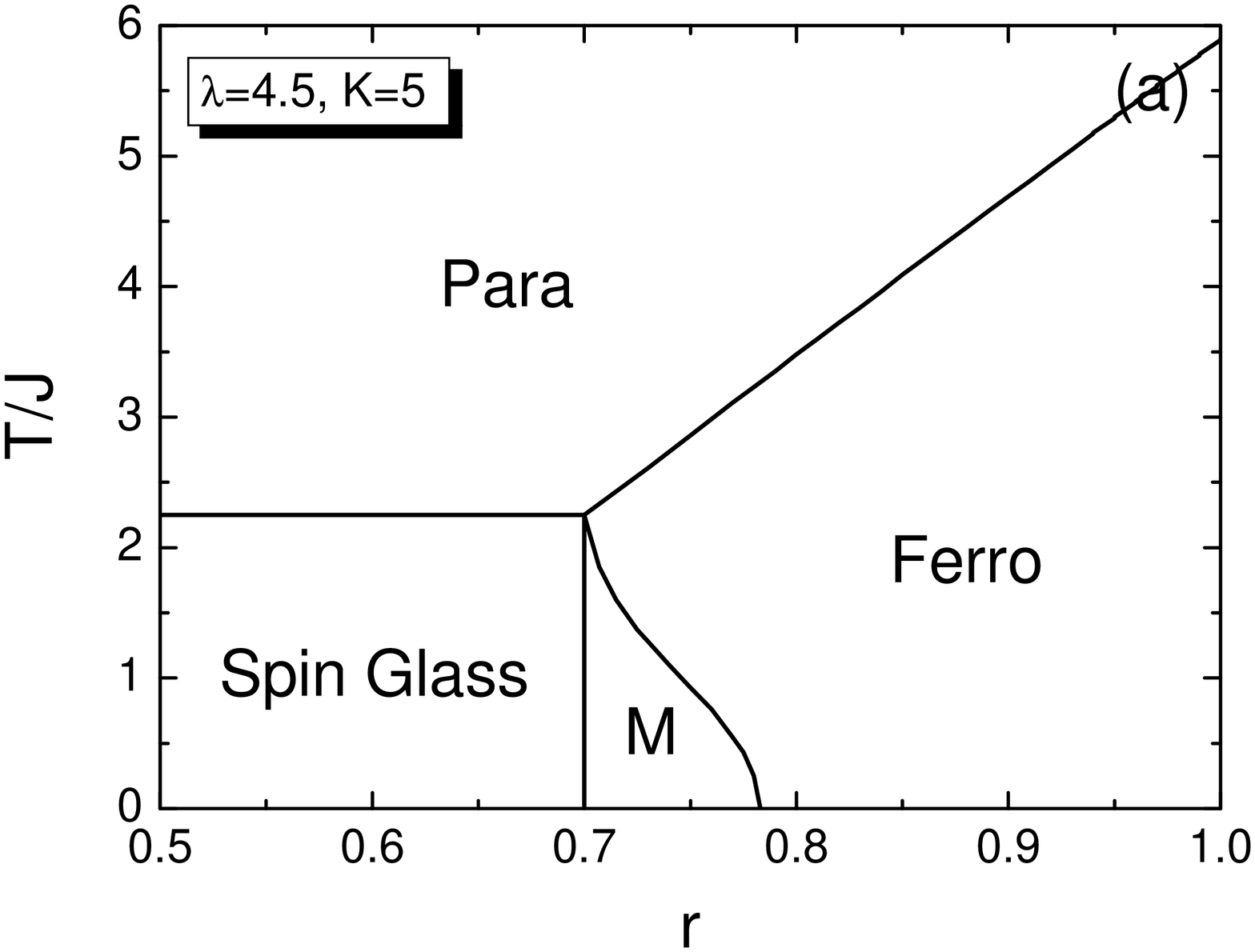}
\includegraphics[angle=270, scale=0.25]{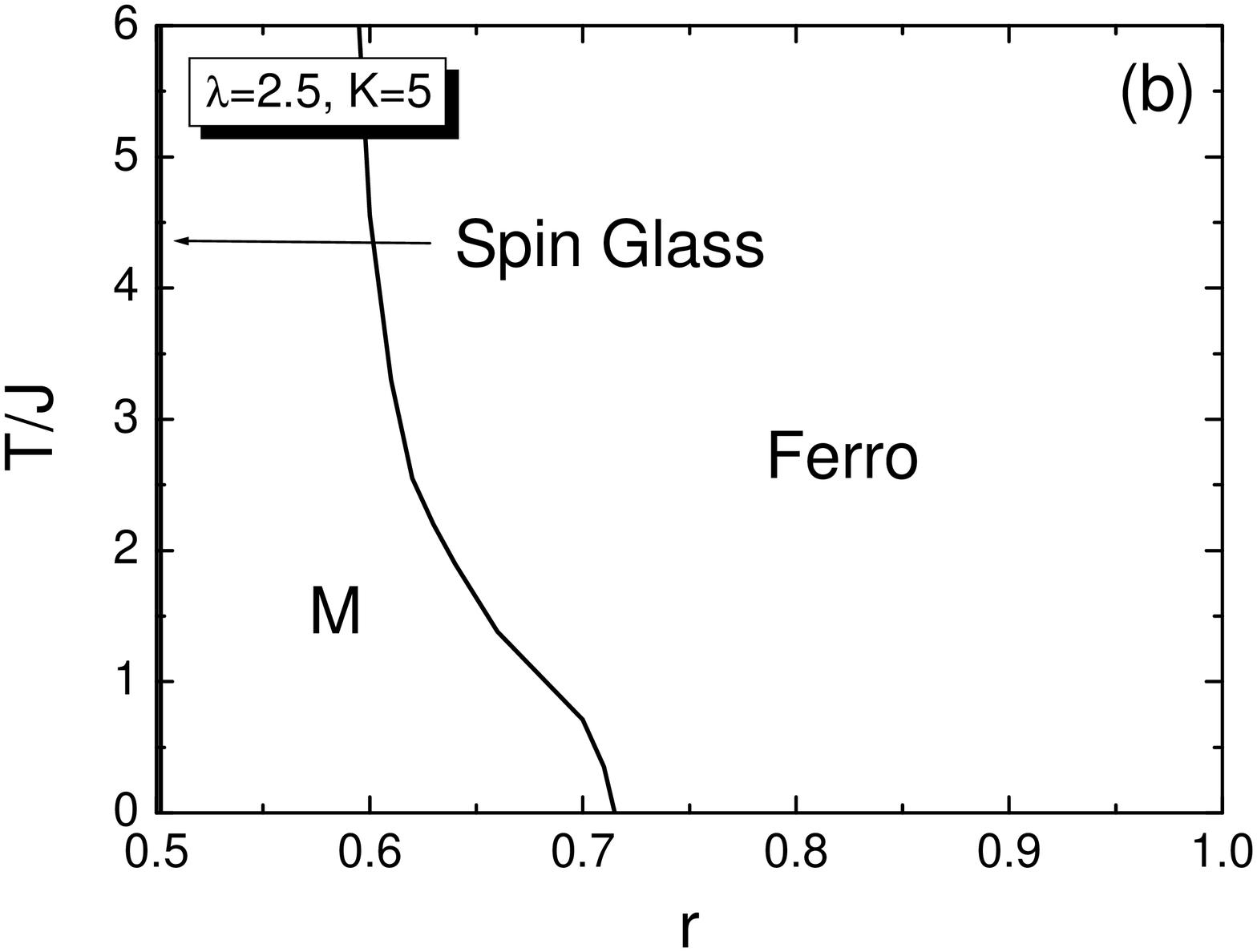}
\caption{The phase diagram in the $(r,T/J)$ plane for $K=5$ with
$\lambda=4.5 (> 3.0)$ (a) and $\lambda=2.5 (< 3.0)$ (b).}
\label{fig:pd_r}
\end{figure}

We also consider the phase diagram in the $(r,T/J)$ plane for
given $\lambda$ and $K$ in FIG.\ref{fig:pd_r}. The phase diagram
is schematically similar to the one for the SK model. In the
original paper of the SK model~\cite{Sherrington75}, a new
coupling constant $J_0$ of the F interactions was introduced and
the ratio $J_0/J$ plays a similar role of the parameter $r$ here.
Accordingly, the phase diagram in the $(r,T/J)$ plane here
corresponds to the one in the $(J_0/J,T/J)$ plane in the work of
the SK model. FIG.\ref{fig:pd_r}(a) shows the phase diagram for
$\lambda > 3$. The formulae of the phase boundaries of P-SG and
P-F are easily derived from Eqs.(\ref{t_c}) and (\ref{t_g}). The
P-SG phase boundary is constant as $1/\tanh^{-1}\sqrt{K_p/K}$,
independent of the parameter $r$ and the P-F phase boundary is
determined as $T/J=1/\tanh^{-1}(K_p/K(2r-1))$. The multicritical
point is determined as
\begin{equation}
\Big(r^{*},\frac{T^{*}}{J}\Big)=\Big(\frac{\sqrt{K_{p}/K}+1}{2},
\frac{1}{\tanh^{-1}\sqrt{K_p/K}}\Big).
\label{t_multi_r}
\end{equation}
The SG-M phase boundary is given by the
vertical line as before. The F-M boundary is obtained from
Eq.(\ref{at}), finding numerically that the region of the M phase
shrinks as $\lambda$ increases, and eventually it remains on the
line spanning from the multicritical point to $T=0$ for a given
$K$, while it exhibits a fat-tail behavior in the direction of the
parameter $K$.

We plot the phase diagram in the $(r,T/J)$ plane
for $\lambda < 3$ with a given $K (>1)$ in FIG.\ref{fig:pd_r}(b).
Note that as $\lambda \to 3$ for a given $K$, $r^*$ approaches
1/2, while $T^*/J$ diverges to infinity. Thus, for $2 < \lambda
<3$, the SG phase can exist only when $r=1/2$.
For $1/2< r < 1$, the F and M phases exist and the F-M boundary
is given by the AT line (Eq.(\ref{at})).

\subsection{The SG order parameter}

In the SG phase $(m = 0, q \neq 0)$, the SG order parameter $q$ is
determined by
\begin{eqnarray}
q = \int {\cal D}z \sum_{i=1}^N p_i \tanh^2 (z\sqrt{NK{\bf T_2}p_i
q}). \label{qa}
\end{eqnarray}
Note that Eq.(\ref{qa}) is independent of $r$ but valid for $1/2
\leq r < r^*$, $r^*$ being the value of $r$ at the multicritical
point.

\begin{figure}
\includegraphics[angle=270, scale=0.35]{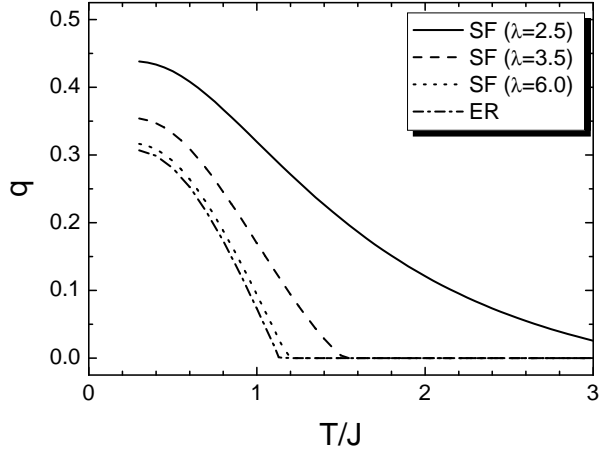}
\caption{The behavior of $q$ in Eq.(\ref{qa}) for $N=5000$ and
$K=2$ for $1/2 \leq r < r^*$.} \label{q_behavior}
\end{figure}

In this section, we determine the critical behavior of $q$ near
the SG transition. The right hand side of Eq.(\ref{qa}) involves a sum
of the type
\begin{eqnarray}
S(y) = \frac{1}{N} \sum_{i=1}^{N} F (Np_i y/(1-\mu))
\end{eqnarray}
with $y=(1-\mu)K{\bf T_2}qz^{2}$ and $F(x)=x
\tanh^{2} \sqrt{x}$. When $y$ is small in $S(y)$, a singular term
$y^{\lambda-1}$ competes with other regular terms. General expressions
for small $y$ expansions are derived in APPENDIX B.
When Eq.(\ref{s_y3}) is used and the Gaussian integration over $z$
is performed, Eq.(\ref{qa}) becomes
\begin{eqnarray}
q/(1-\mu) = \frac{\lambda-1}{\sqrt{\pi}} 2^{\lambda-1} \Gamma
(\lambda - \frac{3}{2}) D(\lambda) Q'^{\lambda-2} -
\frac{\lambda-1}{3-\lambda}Q' + 2
\frac{\lambda-1}{4-\lambda}Q'^{2} + \mathcal{O}(Q'^{3}) \label{qb}
\end{eqnarray}
where
\begin{eqnarray}
D(\lambda) \equiv \left \{
\begin{array}{lllll}
\int_{0}^{\infty} dx~ x^{3-2\lambda} \tanh^{2} x & &
\hbox{for} & &  2 < \lambda < 3, \\
-\int_{0}^{\infty} dx~ x^{3-2\lambda} (x^2-\tanh^{2} x) & &
\hbox{for} & & 3 < \lambda < 4, \\
\end{array}
\right.
\end{eqnarray}
and $Q'=(1-\mu)K{\bf T_2}q=(\lambda-2)K{\bf T_2}q/(\lambda-1)$.
Equating the right hand side of Eq.(\ref{qb}) with
$Q'/(1-\mu)^{2}K{\bf T_2}$, one sees that $Q'^{3-\lambda} \sim
K{\bf T_2} \sim T^{-2}$ for $2 < \lambda <3$, $(1/K_{p} - 1/K{\bf
T_2}) \sim Q'^{\lambda-3}$ for $3< \lambda <4$, and $(1/K_{p} -
1/K{\bf T_2}) \sim Q'$ for $\lambda >4$. Here $K_{p}$ is given by
Eq.(\ref{K_p}) and the $\lambda$-dependent positive coefficients
are neglected. Therefore, as $T \to \infty$ $(2 < \lambda < 3)$ or
$\epsilon_g \equiv (T_g-T)/T_g \to 0$ $(\lambda > 3)$, $q$ behaves
as
\begin{eqnarray}
q \sim \left \{
\begin{array}{lllll}
T^{-2(\lambda-2)/(3-\lambda)} && {\rm for} && 2 < \lambda < 3, \\
\epsilon_g^{1/(\lambda-3)} && {\rm for} && 3 < \lambda < 4, \\
\epsilon_g && {\rm for} && \lambda > 4. \\
\end{array}
\right.
\label{K_p1}
\end{eqnarray}
When $\lambda=3$, use of Eq.(\ref{s_y4}) yields
\begin{eqnarray}
q \sim T^2 \exp(-2T^2/KJ^2) ~~~~ {\rm as} ~~ T \to \infty,
\end{eqnarray}
while, when $\lambda=4$,
\begin{eqnarray}
q \sim \epsilon_g/\ln \epsilon_g^{-1} ~~~~ {\rm as} ~~ \epsilon_g \to 0.
\end{eqnarray}
For general temperatures, $q$ can be obtained numerically from
Eq.(\ref{qa}). The behavior of $q$ for various $\lambda$ are
shown in FIG.\ref{q_behavior}.

\section{The perturbative approach}

In this section, we use the perturbative approach to evaluate the
free energy and to obtain the order parameter behaviors near the
transitions.
For simplicity, we use the notations defined through
$Q_{\alpha} \equiv K \mathbf{T}_{1} q_{\alpha}$, $Q_{\alpha \beta}
\equiv K \mathbf{T}_{2} q_{\alpha \beta}$, $Q_{\alpha \beta
\gamma} \equiv K \mathbf{T}_{3} q_{\alpha \beta \gamma}$,
$Q_{\alpha \beta \gamma \delta} \equiv K \mathbf{T}_{4} q_{\alpha
\beta \gamma \delta}$, and so on. Let $\rm R$ represent a
subset of the replica indices $\{1,2,\ldots, n \}$. Then
it is convenient to denote the set
$\{Q_{\alpha}, Q_{\alpha \beta},\ldots \}$ as $\{Q_{\rm
R}\}$. We also write $\sigma_{\rm R} \equiv \prod_{\alpha
\in {\rm R}} s^{\alpha}=\pm 1$. With these notations, Eq.(\ref{x_i})
becomes $X_i=\sum_{\rm R} Np_i Q_{\rm R}\sigma_{\rm R}$ where
the sum is over all subsets of $\{1,2,\ldots,n\}$ except
the null set, and
\begin{equation}
e^{X_i}=\prod_{\rm R} e^{Np_i Q_{\rm R} \sigma_{\rm R}}
=\prod_{\rm R} \cosh Np_iQ_{\rm R} \prod_{\rm R}(1+ \tau_{\rm R}
\sigma_{\rm R})
\end{equation}
with $\tau_{\rm R} \equiv \tanh Np_iQ_{\rm R}$. Our perturbative
approach is to expand $\prod_{\rm R}(1+\tau_{\rm R}\sigma_{\rm
R})$ and keep only the terms up to given order. In the ER limit
$\lambda \to \infty$, we anticipate that $\tau_{\alpha}\sim
\epsilon_c^{1/2}$, $\tau_{\alpha \beta}\sim \epsilon_c$, etc.
from VB~\cite{Viana85}, where $\epsilon_c \equiv (T_c-T)/T_c$
is the reduced temperature.

Using the properties that ${\rm Tr} \sigma_{\rm R}=0$,
${\rm Tr} \sigma_{\rm R}\sigma_{\rm R'}=0$ for ${\rm R}\ne {\rm R'}$
and so on, the first few terms relevant to our discussion below are
\begin{eqnarray}
n\beta f =&& \frac{1}{2 K \mathbf{T}_{1}} \sum_{\alpha}
Q_{\alpha}^{2} + \frac{1}{2 K \mathbf{T}_{2}} \sum_{\alpha <
\beta} Q_{\alpha \beta}^{2} + \frac{1}{2 K \mathbf{T}_{3}}
\sum_{\alpha < \beta < \gamma} Q_{\alpha \beta \gamma}^{2} +
\frac{1}{2 K \mathbf{T}_{4}} \sum_{\alpha < \beta
< \gamma < \delta} Q_{\alpha \beta \gamma \delta}^{2} \nonumber\\
&&-\frac{1}{N}\sum_{i} \sum_{\rm R} \ln \cosh (Np_{i} Q_{\rm R})
-\frac{1}{N}\sum_{i} \Big[ \sum_{\alpha < \beta}
\tau_{\alpha} \tau_{\beta} \tau_{\alpha \beta}
+\sum_{\alpha <\beta <\gamma}\tau_{\alpha} \tau_{\beta} \tau_{\gamma}
\tau_{\alpha \beta \gamma}\nonumber \\
&&+\sum_{\alpha < \beta < \gamma}
(\tau_{\alpha}\tau_{\beta}\tau_{\beta \gamma}\tau_{\alpha \gamma}+
\tau_{\beta}\tau_{\gamma}\tau_{\alpha \beta}\tau_{\alpha\gamma}+
\tau_{\gamma}\tau_{\alpha}\tau_{\alpha\beta}\tau_{\beta \gamma})
+\sum_{\alpha < \beta < \gamma} \tau_{\alpha
\beta} \tau_{\beta \gamma} \tau_{\alpha \gamma} \nonumber \\
&&+\sum_{\alpha <\beta < \gamma < \delta} \tau_{\alpha}\tau_{\beta}
\tau_{\gamma}\tau_{\delta} \tau_{\alpha \beta \gamma \delta}
+\sum_{\alpha < \beta < \gamma < \delta}(
\tau_{\alpha\beta}\tau_{\gamma\delta}+
\tau_{\alpha\gamma}\tau_{\beta\delta}+
\tau_{\alpha\delta}\tau_{\beta\gamma})
\tau_{\alpha\beta\gamma\delta}\nonumber \\
&&+\sum_{\alpha < \beta < \gamma < \delta}\tau_{\alpha\beta}
\tau_{\beta\gamma} \tau_{\gamma\delta}\tau_{\alpha\delta}\Big].
\label{free_energy2}
\end{eqnarray}

The result of APPENDIX B with $F(x)=\ln \cosh x$ gives
\begin{eqnarray}
\frac{1}{N}\sum_{i} \ln \cosh (Np_{i} Q_{\rm R})=
A(\lambda)Q_{\rm R}^{\lambda-1}+\frac{a_2}{2}Q_{\rm
R}^{2} -\frac{a_{4}}{12}Q_{\rm R}^{4}+{\mathcal O}(Q_{\rm R}^6),
\end{eqnarray}
where
\begin{eqnarray}
A(\lambda)\equiv \left \{
\begin{array}{lllll}
\frac{(\lambda-2)^{\lambda-1}}{(\lambda-1)^{\lambda-2}}
\int_{0}^{\infty} dx~ x^{-\lambda} \ln \cosh x & &
\hbox{for} & &  2 < \lambda < 3, \\
\frac{(\lambda-2)^{\lambda-1}}{(\lambda-1)^{\lambda-2}}
\int_{0}^{\infty} dx~ x^{-\lambda}
\Big(\ln \cosh x - \frac{1}{2}x^{2} \Big)
& & \hbox{for} & & 3 < \lambda < 5, \\
\frac{(\lambda-2)^{\lambda-1}}{(\lambda-1)^{\lambda-2}}
\int_{0}^{\infty} dx~ x^{-\lambda}\Big(\ln \cosh x-
\frac{1}{2}x^{2}+\frac{1}{12}x^{4} \Big) & &
\hbox{for} & & 5 < \lambda < 7,
\end{array}
\right.
\end{eqnarray}
and
\begin{eqnarray}
a_l=(\lambda-2)^{l}/[(\lambda-1)^{l-1}(\lambda-1-l)] \label{a_l}.
\end{eqnarray}
The last sums in Eq.(\ref{free_energy2}) can
be represented as integrals as
\begin{eqnarray}
\frac{1}{N}\sum_{i} \tau_{\rm R} \tau_{\rm R'} \cdots =
(\lambda-1)\int_{1}^{\infty} dz~ z^{-\lambda} \tanh zQ'_{\rm R}
\tanh zQ'_{\rm R'} \cdots
\label{product}
\end{eqnarray}
with $Q'_{\rm R}\equiv (\lambda-2)Q_{\rm R}/(\lambda-1)$.

\subsection{The replica symmetric free energy}

We derive the RS solution of the order parameters up to the
fourth order with the notations of $Q_{\alpha}=M$, $Q_{\alpha
\beta}=Q$, $Q_{\alpha \beta \gamma}=Q_{3}$ and $Q_{\alpha \beta
\gamma \delta}=Q_{4}$, respectively.
Then the terms in Eq.(\ref{product}) take the form of
\begin{eqnarray}
\mathcal{B}_{n_1,n_2,n_3,n_4} \equiv
(\lambda-1)\int_{1}^{\infty} dz~ z^{-\lambda}
\tanh^{n_{1}} zM' \tanh^{n_2} zQ' \tanh^{n_3} zQ_3'
\tanh^{n_4} zQ'_{4},
\label{b_def}
\end{eqnarray}
where $M' \equiv (1-\mu)M=(\lambda-2)M/(\lambda-1)$,
$n_1,\ldots,n_4$ are integers, and other primed quantities are
similarly defined.

The RS free energy $f(M,Q,Q_3,Q_4)$ in the limit of $n \to 0$
is then written as
\begin{eqnarray}
{\beta f} &=& \frac{b_1}{2} M^{2} -
\frac{b_2}{4} Q^{2} + \frac{b_3}{6} Q_{3}^{2} -
\frac{b_4}{8} Q_{4}^{2} + \frac{a_{4}}{12} M^{4} -
\frac{a_{4}}{24} Q^{4} + \frac{a_{4}}{36} Q_{3}^{4} -
\frac{a_{4}}{48} Q_{4}^{4} \nonumber\\
&& -A(\lambda)\Big[M^{\lambda-1}-\frac{1}{2}
Q^{\lambda -1}+\frac{1}{3} Q_{3}^{\lambda -1} - \frac{1}{4}
Q_{4}^{\lambda -1}\Big] \\&&+ \frac{1}{2} \mathcal{B}_{2,1,0,0}
- \frac{1}{3} [\mathcal{B}_{3,0,1,0} + 3\mathcal{B}_{2,2,0,0}
+ \mathcal{B}_{0,3,0,0}]+\frac{1}{4} [\mathcal{B}_{4,0,0,1}+
3\mathcal{B}_{0,2,0,1}+\mathcal{B}_{0,4,0,0}],
\nonumber
\end{eqnarray}
where $b_l \equiv (K \mathbf{T}_{l})^{-1}-a_{2}$ for $l=1,2,3$ and
4. Note that $1/a_2$ is nothing but $K_p$ for $\lambda > 3$
given in Eq.(\ref{K_p}), while it is negative for $2 < \lambda < 3$.

The RS solutions of $M$, $Q$, $Q_{3}$ and $Q_{4}$ are
obtained by solving the self-consistent equations,
\begin{eqnarray}
\partial f/\partial M = \partial f/\partial Q =
\partial f/\partial Q_3 = \partial f/\partial Q_{4} = 0.
\label{self_con}
\end{eqnarray}
When $M,Q,Q_3$ and $Q_4$ are small, ${\cal B}_{n_1,\cdots,n_4}$
are small. Their leading order behaviors are calculated in APPENDIX C.

The phase boundary of the P-F transition is determined as the same
obtained in the SK approach. When $2 < \lambda < 3$, since
$A(\lambda)$ is nonzero and positive, the transition
temperature $T_{c}$ becomes infinity so that the system is always
in the F phase when $r > 1/2$. For $r=1/2$, however, $b_1=\infty$,
and $M^2$ has to be zero. Then the system is in the SF phase.

\subsection{The P-F transition and the order parameters}

We first consider the P-F transition. In the F phase, all the four
order parameters remain nonzero. The behaviors of each order parameter
within leading order are discussed below and listed in TABLE I.

\begin{table}[b]
\begin{tabular}{c|ccccc}
\hline
~order parameters~ & ~~~~~$2<\lambda<3$~~~~~ & ~~~~~$3<\lambda<4$~~~~~ & ~~~~~$4<\lambda<5$~~~~~ & ~~~~~$5<\lambda<6$~~~~~ & ~~~~~$\lambda>6$~~~~~ \\
\hline
$M$     & $\sim T^{-1/(3-\lambda)}$   & $\sim \epsilon_{c}^{1/(\lambda-3)}$           & $\sim \epsilon_{c}^{1/(\lambda-3)}$           & $\sim \epsilon_{c}^{1/2}$           & $\sim \epsilon_{c}^{1/2}$ \\
$m$     & $\sim T^{-(\lambda-2)/(3-\lambda)}$   & $\sim \epsilon_{c}^{1/(\lambda-3)}$           & $\sim \epsilon_{c}^{1/(\lambda-3)}$           & $\sim \epsilon_{c}^{1/2}$           & $\sim \epsilon_{c}^{1/2}$ \\
$Q$     & $\sim T^{-(4-\lambda)/(3-\lambda)}$ & $\sim \epsilon_{c}^{(\lambda-2)/(\lambda-3)}$ & $\sim \epsilon_{c}^{2/(\lambda-3)}$           & $\sim \epsilon_{c}^{1}$             & $\sim \epsilon_{c}^{1}$ \\
$q$     & $\sim T^{-(\lambda-2)/(3-\lambda)}$ & $\sim \epsilon_{c}^{(\lambda-2)/(\lambda-3)}$ & $\sim \epsilon_{c}^{2/(\lambda-3)}$           & $\sim \epsilon_{c}^{1}$             & $\sim \epsilon_{c}^{1}$ \\
$Q_{3}$ & $\sim T^{-(7-2\lambda)/(3-\lambda)}$ & $\sim \epsilon_{c}^{(\lambda-2)/(\lambda-3)}$ & $\sim \epsilon_{c}^{(\lambda-2)/(\lambda-3)}$ & $\sim \epsilon_{c}^{3/2}$           & $\sim \epsilon_{c}^{3/2}$ \\
$q_{3}$ & $\sim T^{-(\lambda-2)/(3-\lambda)}$ & $\sim \epsilon_{c}^{(\lambda-2)/(\lambda-3)}$ & $\sim \epsilon_{c}^{(\lambda-2)/(\lambda-3)}$ & $\sim \epsilon_{c}^{3/2}$           & $\sim \epsilon_{c}^{3/2}$ \\
$Q_{4}$ & $\sim T^{-(10-3\lambda)/(3-\lambda)}$ & $\sim \epsilon_{c}^{(\lambda-2)/(\lambda-3)}$ & $\sim \epsilon_{c}^{(\lambda-2)/(\lambda-3)}$ & $\sim \epsilon_{c}^{(\lambda-2)/2}$ & $\sim \epsilon_{c}^{2}$ \\
$q_{4}$ & $\sim T^{-(\lambda-2)/(3-\lambda)}$ & $\sim \epsilon_{c}^{(\lambda-2)/(\lambda-3)}$ & $\sim \epsilon_{c}^{(\lambda-2)/(\lambda-3)}$ & $\sim \epsilon_{c}^{(\lambda-2)/2}$ & $\sim \epsilon_{c}^{2}$ \\
\hline
\end{tabular}
\caption{The $\lambda$-dependent critical behaviors of the four order
parameters and their scaled quantities (Eq.(\ref{scaled_quantities}))
under the P-F transition. Here $\epsilon_c \equiv
(T_c-T)/T_c$ is the reduced temperature.}
\end{table}

\begin{itemize}
\item[(i)] When $2 < \lambda < 3$, the leading order terms
in free energy $\beta f$ are
\begin{eqnarray}
\beta f &\simeq& -A(\lambda) M^{\lambda -1}+
\frac{b_1}{2}M^{2} + \Big(\frac{1}{2} A(\lambda)
-\mathcal{C}_{2,0}-\frac{1}{3}\mathcal{C}_{3,0} +
\frac{1}{4}\mathcal{C}_{4,0}+\cdots\Big) Q^{\lambda -1} \nonumber \\
&&-\frac{b_2}{4}Q^{2}-\frac{1}{3} A(\lambda) Q_{3}^{\lambda
-1} + \frac{b_3}{6}Q_{3}^{2} +
\frac{1}{4} A(\lambda) Q_{4}^{\lambda -1}-
\frac{b_4}{8}Q_{4}^{2} \nonumber \\ && +
\frac{1}{2}\mathcal{C}_{2,1}M^{\lambda-2}Q -
\frac{1}{3}\mathcal{C}_{3,1}M^{\lambda-2}Q_{3} +
\frac{1}{4}\mathcal{C}_{4,1}M^{\lambda-2}Q_{4}
\end{eqnarray}
from TABLE III with $\mathcal{C}_{n,p}$ given in Eq.(\ref{c_6}).

By applying Eq.(\ref{self_con}) to the free energy, we obtain the
self-consistent equations for the four order parameters. Note that
from the definition of $b_{l} \equiv (K{\bf T}_l)^{-1}-a_2$, we
find that $b_l\sim T^l$ as $T\to \infty$. All other coefficients such as
$A(\lambda)$ and $\{ {\cal{C}}_{n,p} \}$ are independent of
$T$. From $\partial f/\partial M =0$, we obtain that
$-(\lambda-1)A(\lambda) M^{\lambda -2}+b_1 M=0$,
leading to that
$M \sim [(\lambda-1)A(\lambda)/b_1]^{1/(3-\lambda)}]\sim
T^{-1/(3-\lambda)}$. From $\partial f/\partial Q=0$, we obtain that
$(A(\lambda)-2{\cal{C}}_{2,0}+\cdots)(\lambda-1)Q^{\lambda-2}+
\mathcal{C}_{2,1} M^{\lambda-2}-b_2 Q=0$. Since the second term is
more dominant than the first, we obtain that $Q \sim
{\cal C}_{2,1}M^{\lambda-2}/b_2 \sim T^{-(4-\lambda)/(3-\lambda)}$.
Fortunately, the coefficient of $Q^{\lambda-1}$ is not needed
to determine the leading order behavior of $Q$. Similarly, we
obtain that $Q_3\sim T^{-(7-2\lambda)/(3-\lambda)}$ and $Q_4\sim
T^{-(10-3\lambda)/(3-\lambda)}$. Subsequently, we obtain
$m\sim q \sim q_3 \sim q_4 \sim T^{-(\lambda-2)/(3-\lambda)}$,
where
\be
m=M/K{\bf T_1},~~q=Q/K{\bf T_2},~~q_3=Q_3/K{\bf T_3},~~\hbox{and}
~~~q_4=Q_4/K{\bf T_4}.
\label{scaled_quantities}
\ee
It is noteworthy that the behavior
of $m$ is different from that of the unweighted magnetization,
${\bar m} \sim T^{-1/(3-\lambda)}$, where ${\bar m}=(1/N)\sum_i
\langle s_i \rangle$ as previously studied in
Ref.\cite{Dorogovtsev02b,Leone02}.
This is because ${\bar m}\sim M$ to the leading order.

\item[(ii)]
When $\lambda > 3$, the transition temperature $T_{c}$
is determined by
\begin{eqnarray}
b_1(T_{c})=0, ~~~ \textrm{i.e.,} ~~~ a_{2}K\mathbf{T}_1(T_c)=1
\end{eqnarray}
which is the same as Eq.(\ref{t_c}). When $3 < \lambda < 4$, the
leading order terms in $\beta f$ are
\begin{eqnarray}
\beta f &\simeq& \frac{b_{1}}{2}M^{2}-A(\lambda)
M^{\lambda -1} - \frac{b_{2}}{4}Q^{2} +
\frac{1}{2}\mathcal{C}_{2,1} M^{\lambda-2}Q \nonumber \\
&&+\frac{b_3}{6}Q_{3}^{2}-\frac{1}{3}\mathcal{C}_{3,1}M^{\lambda-2}Q_{3} -
\frac{b_{4}}{8}Q_{4}^{2} +
\frac{1}{4}\mathcal{C}_{4,1}M^{\lambda-2}Q_{4}.
\end{eqnarray}
Note that $A(\lambda) < 0$ for $3 < \lambda <5$.
The most leading term is $(b_1/2)M^2$ and the transition
temperature $T_c$ is determined by $b_1=0$. Just below $T_c$,
$b_1 <0$ and $|b_1| \sim {\cal O}(\epsilon_c)$,
where $\epsilon_c=(T_c-T)/T_c$. From
$\partial f/\partial M =0$, we obtain that
$-(\lambda-1) A(\lambda) M^{\lambda -2} + b_1 M=0$,
leading to that $M \sim \epsilon_c^{1/(\lambda-3)}$. From
$\partial f/\partial Q=0$, we obtain that $\mathcal{C}_{2,1}
M^{\lambda-2}-b_2 Q=0$. Since $b_2$ is constant near $T_c$,
we obtain that $Q \sim M^{\lambda-2} \sim
{\epsilon_c}^{(\lambda-2)/(\lambda-3)}$.
Similarly, it is obtained that $Q_3 \sim M^{\lambda-2} \sim
{\epsilon_c}^{(\lambda-2)/(\lambda-3)}$ and $Q_4 \sim
M^{\lambda-2} \sim {\epsilon_c}^{(\lambda-2)/(\lambda-3)}$. Unlike
the case of $2 < \lambda < 3$, $m\sim M$, $q\sim Q$, $q_3 \sim
Q_3$, and $q_4 \sim Q_4$. Such relations hold for all $\lambda > 3$.

\item[(iii)] When $4 < \lambda < 5$,
the free energy is written as
\begin{eqnarray}
\beta f &\simeq& \frac{b_{1}}{2}M^{2}- A(\lambda)
M^{\lambda -1}-\frac{b_{2}}{4}Q^{2}+\frac{a_{3}}{2}M^{2}Q
\nonumber\\
&& + \frac{b_{3}}{6}Q_{3}^{2} -
\frac{1}{3}\mathcal{C}_{3,1}M^{\lambda-2}Q_{3} -
\frac{b_{4}}{8}Q_{4}^{2} +
\frac{1}{4}\mathcal{C}_{4,1}M^{\lambda-2}Q_{4}.
\end{eqnarray}
Following the same step as used in $3 < \lambda < 4$, we
obtain that $M \sim \epsilon_c^{1/(\lambda-3)}$,
$Q \sim M^2 \sim \epsilon_c^{2/(\lambda-3)}$ and
$Q_{3} \sim Q_{4} \sim M^{\lambda-2} \sim
\epsilon_c^{(\lambda-2)/(\lambda-3)}$.

\item[(iv)] When $5 < \lambda < 6$, the free energy is written as
\begin{eqnarray}
\beta f \simeq && \frac{b_{1}}{2}M^{2} + \frac{a_{4}}{12}M^{4} -
\frac{b_{2}}{4}Q^{2} + \frac{a_{3}}{2}M^{2}Q\nonumber \\
&& +\frac{b_{3}}{6}Q_{3}^{2}-\frac{a_{4}}{3}M^{3}Q_{3} -
\frac{b_{4}}{8}Q_{4}^{2}+\frac{1}{4}\mathcal{C}_{4,1}M^{\lambda-2}Q_{4}.
\end{eqnarray}
Following the same steps as before, we obtain that
$M \sim \epsilon_c^{1/2}$, $Q \sim \epsilon_c$, $Q_{3}
\sim \epsilon_c^{3/2}$, and $Q_{4} \sim \epsilon_c^{(\lambda-2)/2}$.

\item[(v)] When $\lambda > 6$, the free energy is written as
\begin{eqnarray}
\beta f \simeq \frac{b_1}{2}M^{2}+\frac{a_4}{12}M^{4} -
\frac{b_2}{4}Q^{2} + \frac{a_3}{2}M^{2}Q +
\frac{b_3}{6}Q_{3}^{2} - \frac{a_4}{3}M^{3}Q_{3} -
\frac{b_4}{8}Q_{4}^{2} + \frac{a_5}{4}M^{4}Q_{4}.
\end{eqnarray}
Using the same step as before, it is obtained that $M \sim
\epsilon_c^{1/2}$, $Q \sim \epsilon_c$, $Q_{3} \sim
\epsilon_c^{3/2}$ and $Q_{4} \sim \epsilon_c^2$.
\end{itemize}
It is interesting to note that as $\lambda$ increases,
the order parameters progressively acquire the classical mean
field behavior $Q_n\sim \epsilon_c^{n/2}$ starting from the lower
order ones.

\subsection{The P-SG transition and the order parameters}

Here we consider the P-SG transition. In the SG phase, $M$ and
$Q_3$ are always zero for all temperatures. Thus, the free energy
becomes simpler compared with that in the F phase. Using the same
method as used in the P-F transition, we obtain the P-SG
transition temperature and the order parameters $Q$ and $Q_4$ in
various region of $\lambda$, which is listed in TABLE II.

\begin{table}[b]
\begin{tabular}{c|ccc}
\hline
~order parameters~ & ~~~~~$2<\lambda<3$~~~~~~~~~~~~~~ & ~~~~~$3<\lambda<4$~~~~~ & ~~~~~$\lambda>4$~~~~~ \\
\hline
$Q$~~~& $\sim T^{-2/(3-\lambda)}$~~~&
$\sim \epsilon_g^{1/(\lambda-3)}$ & $\sim \epsilon_g^{1}$ \\
$q$~~~& $\sim T^{-2(\lambda-2)/(3-\lambda)}$ &
$\sim \epsilon_g^{1/(\lambda-3)}$ & $\sim \epsilon_g^{1}$ \\
$Q_{4}$~~~& $\sim T^{-(8-2\lambda)/(3-\lambda)}$~&
$\sim \epsilon_g^{(\lambda-2)/(\lambda-3)}$ & $\sim \epsilon_g^{2}$ \\
$q_{4}$~~~& $T^{-2(2-\lambda)/(3-\lambda)}$ &
$\sim \epsilon_g^{(\lambda-2)/(\lambda-3)}$ & $\sim \epsilon_g^{2}$ \\
\hline
\end{tabular}
\caption{The $\lambda$-dependent behaviors of the two order
parameters and their scaled quantities in Eq.(\ref{scaled_quantities})
under the P-SG transition, where $\epsilon_g \equiv (T_g-T)/T_g$.}
\end{table}

For more details, we first determine the P-SG phase boundary. When
$2 < \lambda < 3$, since $A(\lambda)$, the
coefficient of $Q^{\lambda-1}$ is nonzero for all $T$, the spin
glass transition temperature $T_{g}$ is infinity, and no P phase
exists for all $T$. When $\lambda > 3$, the transition point
$T_{g}$ is determined by the formula
\begin{eqnarray}
b_{2}(T_{g}) = 0, ~~\textrm{i.e.,}~~ a_{2}K\mathbf{T}_{2}(T_{g})= 1,
~~{\rm or}~~ K{\bf T}_2(T_g)=K_p
\end{eqnarray}
which is the same as derived in the SK method.
In the SG phase, the order parameter behaves as follows:

\begin{itemize}
\item[(i)] When $2 < \lambda < 3$, the leading order terms
of $\beta f$ read off from TABLE III with $M=Q_3=0$ are
\begin{eqnarray}
\beta f \simeq && \Big(\frac{1}{2}A(\lambda)
-\frac{1}{3}\mathcal{C}_{3,0}+
\frac{1}{4}\mathcal{C}_{4,0}+\cdots \Big) Q^{\lambda -1} -
\frac{b_{2}}{4}Q^{2}\nonumber \\
&&+\frac{1}{4}\Big(A(\lambda)Q_{4}^{\lambda -1} -
\frac{b_{4}}{2}Q_{4}^{2} + 3\mathcal{C}_{2,1}Q^{\lambda-2}Q_{4}\Big).
\end{eqnarray}
By applying $\partial f/\partial Q=\partial f/\partial Q_4=0$, we
obtain that $Q \sim T^{-2/(3-\lambda)}$ and $Q_4 \sim
Q^{\lambda-2}/T^4 \sim T^{-(8-2\lambda)/(3-\lambda)}$. Using the
relation $Q=K{\bf T}_2 q$ and $Q_4=K{\bf T}_4 q_4$, we obtain that
$q\sim q_4 \sim T^{-2(\lambda-2)/(3-\lambda)}$. The result of $q$
is the same as the one derived through the SK method,
Eq.(\ref{K_p1}). Note that the coefficient of $Q^{\lambda-1}$ in
the perturbative approach is in the form of infinite series while
the same is obtained in a closed form in Eq.({\ref{qb}}).

\item[(ii)] When $3 < \lambda < 4$, the free energy is
\begin{eqnarray}
\beta f \simeq -\frac{b_2}{4}Q^{2} + [A(\lambda)/2 -
\mathcal{C}_{3,0}/3 + \mathcal{C}_{4,0}/4] Q^{\lambda -1} -
\frac{b_4}{8}Q_{4}^{2}+
\frac{3}{4}\mathcal{C}_{2,1}Q^{\lambda-2}Q_{4}.
\end{eqnarray}
We note that the coefficient $b_2 \sim -\epsilon_g$ with
$\epsilon_g \equiv (T_g-T)/T_g$. Then we obtain $Q \sim
\epsilon_g^{1/(\lambda-3)}$ Similarly, from $\partial f/\partial
Q_4=0$, we obtain $Q_{4} \sim Q^{\lambda-2} \sim
\epsilon_g^{(\lambda-2)/(\lambda-3)}$ with $b_4$ being constant.

\item[(iii)] When $\lambda > 4$, we have
\begin{eqnarray}
\beta f \simeq -\frac{b_2}{4}Q^{2} - \frac{a_3}{3}Q^{3} -
\frac{b_4}{8}Q_{4}^{2} + \frac{3}{4}a_{3}Q^{2}Q_{4}.
\end{eqnarray}
By following the same step above, we obtain that
$Q \sim \epsilon_g$ and $Q_{4} \sim \epsilon_g^2$.

\end{itemize}

\section{conclusions}

We have studied the spin glass phase transition on SF networks
through the static model. The model contains generic
vertex-weights in it, and edges between two vertices are connected
with the probability given in Eqs. (\ref{f_ij}) and (\ref{p_i}).
The static model enables one to study the spin glass problem using
the replica method by generalizing the dilute Ising spin glass
model with infinite-range interactions. Here we obtained the
replica-symmetric solutions through the two methods, the
Sherrington-Kirkpatrick approach and the perturbative approach. We
also found the phase diagram consisting of the paramagnetic (P),
ferromagnetic (F), spin glass (SG), and mixed (M) phases in the
space of temperature $T$, the mean degree $K$, the fraction of the
ferromagnetic interactions $r$, and the degree exponent $\lambda$.
The AT line was also obtained numerically. The phase diagram is
shown in the $(K,T)$ and $(r,T)$ planes, which are presented in
Figs.~1 and 2, respectively. The critical temperatures $T_c$ and
$T_g$ for the P-F and P-SG phase transitions are simply related to
the percolation threshold $K_p$ in Eqs.(\ref{t_c}) and
(\ref{t_g}). We obtain the same results in the two approaches.
Thus $T_c$ and $T_g$ are infinite when $2 < \lambda \le 3$. The
magnetization and the spin glass order parameter are modified to
account for the inhomogeneity of vertex degrees as $m=\sum_i p_i
\langle s_i^{\alpha} \rangle_i$ and $q=\sum_i p_i \langle
s_i^{\alpha} s_i^{\beta} \rangle_i$, where $p_i$ is the weight of
vertex $i$. Such quantities depend on the degree exponent
$\lambda$. When $2 < \lambda < 3$, due to the fact that
$T_c=\infty$ and $T_g=\infty$, $m$ and $q$ decay as power-laws for
large $T$ as shown in TABLES I and II, which is different from the
patterns of $\bar m$ and $\bar q$, defined with $p_i=1/N$. When
$\lambda > 3$, the order parameters exhibit continuous phase
transitions across $T_c$ and $T_g$, and the associated exponents
depend on $\lambda$, which are listed in TABLES I and II. As
$Q_3$, $Q_4, \ldots$ are of higher orders, the SK approach in Sec.
III, and the perturbative one in Sec. IV give the identical
results for $m$ and $q$ to the leading order. We find the critical
exponents for the P/SG transition are non-classical in the range
$3 < \lambda < 4$, corresponding to $3 < \lambda < 5$ for the P/F
one~\cite{Dorogovtsev02b}. We have not presented our results at
integer values of $\lambda$ in Section IV for simplicity. At the
borderline cases of $\lambda$, the logarithmic corrections as
given in Eqs.~(\ref{s_y4}), (\ref{c_3}) and (\ref{c_7}) should be considered
explicitly. We mention that the finite-size effect is an important
issue especially for $2 < \lambda \le 3$ which we leave for a
further study.

It is noteworthy that the method we developed here can be applied to
other problems in equilibrium statistical physics on SF networks.
A novelty in this approach is that one needs not rely on the local treelike
structure of SF networks used {\em e.g.} in \cite{Dorogovtsev02b}. 
The result of the phase diagram and the behavior of the order
parameters may be helpful in understanding emerging patterns in
various systems with competing interactions such as social or
biological systems. 
For example, in the region  $2<\lambda \le 3$
where most real-world SF networks belong, it is known that the structural 
characteristic of the network is so dominant that homogeneously interacting 
systems are in the ordered state for all temperatures. Our result shows
that 
it is also the case even when there are competing interactions. Also for  
$2<\lambda \le 3$, the fact that a slight dominance of cooperative
interactions
 ($r \gtrsim 1/2$) drives the system into the ferromagnetically ordered 
or the mixed state
suggests that most social and biological systems would be driven into 
the majority state (ferromagnetic or mixed state) at equilibrium. While the
current 
study is meaningful as a first step of understanding thermodynamic
property 
of the systems with competing interactions, further studies have to be
followed 
towards real-world systems where the signs of interactions may be 
correlated with the degrees of vertices, or the interaction signs may
change
with time as in the prisoner's dilemma problem. 

While preparing this manuscript, we have learned of a recent preprint
by Mooij and Kappen~\cite{Mooij04}, which addressed the same issue.
They used the Bethe approximation to obtain a criterion
for $T_g$ and applied it to the $\lambda=\infty$ and $\lambda=3$
cases numerically. Our work gives analytic results for $T_g$
as well as physical ones such as the phase diagram and the behaviors
of the order parameters, which depend on the degree exponent.

\begin{acknowledgments}
This work is supported by the KOSEF Grant No. R14-2002-059-01000-0
in the ABRL program, and by the Royal Society, London. We thank
J.W. Lee and K.-I. Goh for helpful discussions.
\end{acknowledgments}

\appendix
\section{Evaluation of the remainder}
In this APPENDIX A, we show that
\be
\sum_{i<j} \ln (1+f_{ij}S_{ij}) = \sum_{i<j}NKp_ip_jS_{ij} +R
\label{a1}
\ee
with $R < {\cal{O}}(N^{3-\lambda} \ln N)$ for $2<\lambda <3$,
$R < {\cal{O}}((\ln N)^2)$ for $\lambda=3$ and $R < {\cal{O}}(1)$
for $\lambda>3$. Here $S_{ij}=\langle \exp(\beta J_{ij}\sum_{\alpha=1}^n
s_i^\alpha s_j^\alpha) -1 \rangle _r $ is a quantity independent
of the system size $N$. To do so, we expand the logarithm on
the left hand side of Eq.(\ref{a1}) to write it as
\be \sum_{i<j} \ln (1+f_{ij}S_{ij}) = \sum_{i<j}NKp_ip_jS_{ij} +
\sum_{i<j}(f_{ij}-NKp_ip_j)S_{ij}+\sum_{n=2}^\infty \frac{(-1)^{n+1}}{n}
\sum_{i<j}f_{ij}^n S_{ij}^n \label{a2} \ee
and show that the positive quantities defined by
\be
R' \equiv |\sum_{i<j}(NKp_ip_j-f_{ij})S_{ij}| \label{rp1} \ee
and
\be R_n \equiv |\sum_{i<j}f_{ij}^n S_{ij}^n| \label{rn1} \ee
($n \ge 2$) are all bounded above by ${\sl {o}}(N)$ quantities.

First let us consider $R'$. Since $S_{ij}$ are independent
of $N$, we replace $S_{ij}$ by their maximum value
$S_{\rm max} \equiv \max_{i<j}|S_{ij}|$ to get
\be
R' \le S_{\rm max} \sum_{i<j} G_1 (NKp_ip_j) \le \frac{S_{\rm max}}{2}
[{\sum_{i,j} G_1(NKp_ip_j)-G_1(NKp_1^2)}], \label{rp2}
\ee
where
\be G_1(x) \equiv x-1+e^{-x} \label{g1}. \ee
Here we have added $i=j$ terms for $i \ge 2$ on the right hand side of
Eq.(\ref{rp2}) for convenience. Since $G_1(x)$ is monotone increasing
for $x>0$, the summands in Eq.(\ref{rp2}) decrease
as $i$ and $j$ increase.

We utilize the fact that, for a monotone decreasing continuous
function $F(x)$,
a finite sum is bounded above by an integral as
\be
\sum_{i=1}^N F(i) \le \int_1^N F(x)dx + F(1) \label{sumtoint}.
\ee
Applying Eq.(\ref{sumtoint}) twice to Eq.(\ref{rp2}) and using
$p_i = i^{-\mu}/\zeta_N(\mu)$, we have
\be
R' \le  \frac{ S_{\rm max}}{2} \{ \int_1^N \int_1^N
G_1(\frac{NK}{\zeta_N(\mu)^2}
 x^{-\mu}y^{-\mu})dxdy +2 \int_1^N G_1(\frac{NK}{\zeta_N(\mu)^2}
 x^{-\mu})dx \}.
\label{rp3}
\ee

The double integral in the bracket of Eq.(\ref{rp3}) is, by change of
variables,
\be I_1 \equiv (\lambda-1)^2 (N \epsilon^{\lambda-1})^2
 \int_\epsilon^{\epsilon N^\mu}
 \int_\epsilon^{\epsilon N^\mu} \frac{G_1(uv)}{(uv)^\lambda} dudv,
  \label{I1} \ee
with $\lambda=1+1/\mu$ and $\epsilon=\sqrt{K} N^{1/2-\mu}/\zeta_N (\mu)
 \sim {\cal{O}}(N^{-1/2})$.
Note that in Eq.(\ref{I1}) the upper limit of the integrals is
$\epsilon N^{\mu} \sim {\cal{O}}(N^{(3-\lambda)/2(\lambda-1)})$
and the front factor scales as ${\cal{O}}(N^{3-\lambda})$. We
consider the three cases of $\lambda$ separately.

\begin{itemize}
\item[(i)] When $2<\lambda<3$, since $G_1(x) \sim x$ as $x
\rightarrow \infty$ and $\sim x^2$ as $x \rightarrow 0$, the lower
(upper) limit of the double integral in Eq.(\ref{I1}) can be
expended to 0 ($\infty$) to give a finite value and hence
\be I_1
\le (\lambda-1)^2(N \epsilon^{\lambda-1})^2 \int_0^\infty
\int_0^\infty \frac{G_1(uv)}{(uv)^\lambda} dudv \sim
{\cal{O}}(N^{3-\lambda}).
\ee
\item[(ii)] When $\lambda =3$, the
upper limit of the double integral is ${\cal{O}}(1)$ and the
integrand near the lower limit behaves as $\sim (uv)^{-1}$. We use
$0< G_1(x) < x^2 /2$ for $x>0$ to get \be I_1 \le \frac{1}{2}(N
\epsilon^{\lambda-1})^2 (\ln N)^2 \sim {\cal{O}}((\ln N)^2). \ee
\item[(iii)] When $\lambda >3$, proceeding as in the case of (ii),
we find
\be I_1 \le \frac{1}{2} (\lambda-1)^2 \Big(\frac{N
\epsilon^2}{\lambda-3} \Big)^2 \sim {\cal{O}}(1).
\ee
\end{itemize}

The single integral in the bracket of Eq.(\ref{rp3}) is, by change of
variables,
\be I_2 \equiv 2(\lambda-1) N \delta^{\lambda-1}
 \int_\delta^{\delta N^\mu} \frac{G_1(u)}{u^\lambda} du,
  \label{I2} \ee with
  $ \delta = K N^{1-\mu}/\zeta_N^2(\mu)
 \sim {\cal{O}}(N^{\mu-1})$. Note that in Eq.(\ref{I2})
the upper limit of the integrals is
$\delta N^{\mu} \sim {\cal{O}}(N^{(3-\lambda)/(\lambda-1)})$ and the
front factor scales as ${\cal{O}}(N^{3-\lambda})$. We proceed exactly
the same as in the case of the double integral and find that

\begin{itemize}
\item[(i)] When $2<\lambda<3$,~~~ $I_2 \le 2 (\lambda-1) N \delta^{\lambda-1}
 \int_0^\infty \frac{G_1(u)}{u^\lambda} du \sim {\cal{O}}(N^{3-\lambda}) $.
\item[(ii)] When $\lambda=3$,~~~ $I_2 \le N \delta^{\lambda-1} \ln N \sim
             {\cal{O}}(\ln N) $.
\item[(iii)] When $\lambda >3$,~~~
$ I_2 \le \frac{\lambda-1}{\lambda-3} N\delta^2 \sim
{\cal{O}}(N^{-(\lambda-3)/(\lambda-1)}).$
\end{itemize}
Collecting these, we see that $R'$ is bounded above as
\be
R'\le \left\{\begin{array}{ll}
 {\cal{O}}(N^{3-\lambda}) & \mbox{if $2<\lambda<3$}, \\
 {\cal{O}}((\ln N)^2) & \mbox{if $\lambda=3$},   \\
 {\cal{O}}(1) & \mbox{if $\lambda>3$}.
\end{array}\right.
\label{rp4}
\ee

Next we consider $R_n$ with $n \ge 2$. Similarly to Eq.(\ref{rp2}),
we have
\be R_n \le S_{\rm max}^n \sum_{i<j} f_{ij}^n \le
            \frac{ S_{\rm max}^n}{2} (\sum_{i,j} f_{ij}^n -f_{11}^n).
      \label{rn2} \ee
Applying Eq.(\ref{sumtoint}) twice to Eq.(\ref{rn2}),
\be
R_n \le  \frac{ S_{\rm max}^n}{2} \{ \int_1^N \int_1^N
[G_0(\frac{NK}{\zeta_N(\mu)^2}
x^{-\mu}y^{-\mu})]^n dxdy +2 \int_1^N [G_0(\frac{NK}{\zeta_N(\mu)^2}
x^{-\mu})]^n dx \},
\label{rn3}
\ee
where $G_0(x) \equiv 1- e^{-x}$.
At this point, we use the piecewise linear upper bound for $G_0(x)$ by
\be
\tilde{G}_0 \equiv \left\{\begin{array}{ll} x &~~~ \mbox{for $0<x\le 1$}, \\
1 &~~~ \mbox{for $x>1$}. \end{array}\right.
\ee
Since $G_0(x) \le \tilde{G}_0(x)$ for $x>0$, we can write
Eq.(\ref{rn3}) as
\be R_n \le  \frac{ S_{\rm max}^n}{2} \{ (\lambda-1)^2 (N \epsilon^{
 \lambda-1})^2 \int_\epsilon^{\epsilon N^\mu}
\int_\epsilon^{\epsilon N^\mu}
\frac{[\tilde{G}_0(uv)]^n}{(uv)^\lambda} du dv
+2 (\lambda-1)N \delta^{\lambda-1} \int_\delta^{\delta N^\mu}
\frac{[\tilde{G}_0(u)]^n}{u^\lambda} du  \},
\label{rn4}
\ee
where $\epsilon$ and $\delta$ are defined above. Now the integrations in
Eq.(\ref{rn4}) are elementary. Focusing only on the $N$-dependences,
we find that
\be R_n \le \left\{
\begin{array}{ll}
{\cal{O}}(N^{3-\lambda} \ln N) & \mbox{for $2<\lambda<3$}, \\
{\cal{O}}((\ln N)^2 )         & \mbox{for $\lambda=3$ and $n=2$},\\
{\cal{O}}(1)  & \mbox{for $\lambda=3$ and $n \ge 3$},\\
{\cal{O}}(N^{2-n})  & \mbox{for $\lambda>3$ and $2 \le n < \lambda-1$},\\
{\cal{O}}((\ln N)^2 N^{2-n}) & \mbox{for $\lambda>3$ and $ n = \lambda-1$},\\
{\cal{O}}(N^{-n(\lambda-3)/(\lambda-1)}) & \mbox{for $\lambda>3$
and $ n > \lambda-1$}.
\end{array}\right.
\label{rn5}
\ee

Putting these together, we finally have
\be
|R| \le R' + \sum_{n=2}^\infty \frac{R_n}{n}
\le \left\{\begin{array}{ll}
{\cal{O}}(N^{3-\lambda} \ln N) & \mbox{for $2<\lambda<3$}, \\
{\cal{O}}((\ln N)^2 ) & \mbox{for $\lambda=3$},\\
{\cal{O}}(1) & \mbox{for $\lambda>3$}.
\end{array}\right.
\label{rfinal}
\ee

\section{Evaluation of finite sum in general form}

In this APPENDIX B, we derive a general expansion formula for the
sum
\begin{equation}
S(y)=\frac{1}{N}\sum_{i=1}^N F(Np_i y/(1-\mu))
\end{equation}
for small $y (>0)$ and $N\to \infty$ with $p_i=i^{-\mu}/\zeta_N$
and $\lambda=1+1/\mu >2$ as before. We take $F(x)$ to be a
positive monotone increasing function which diverges slower than
$x^{1/\mu}$ as $x\to \infty$ and has an expansion
$F(x)=\sum_{n=0}^{\infty}f_n x^n$. Converting the sum into an
integral as in APPENDIX A, $S(y)$ becomes, in the $N\to \infty$
limit,
\begin{equation}
S(y)=(\lambda-1)y^{\lambda-1}\int_y^{\infty} \frac{F(x)}
{x^{\lambda}}dx.
\label{s_y}
\end{equation}
We first let $\lambda \ne$ integer and $m_0 < \lambda < m_0+1$
for some integer $m_0$. Then we define
\be
{\tilde F}(x)=F(x)-\sum_{n=0}^{m_0-1} f_n x^n
\ee
and divide $F(x)$ into two parts
\begin{equation}
F(x)=\sum_{n=0}^{m_0-1} f_n x^n+{\tilde F}(x),
\label{f_x}
\end{equation}
Plugging Eq.(\ref{f_x}) into Eq.({\ref{s_y}), the first finite sum
can be integrated term by term to give
\begin{equation}
S(y)=(\lambda-1)\sum_{n=0}^{m_0-1}\frac{f_n}{\lambda-n-1}y^n
+(\lambda-1)y^{\lambda-1}\Big[\int_0^{\infty}\frac{\tilde F(x)}{x^{\lambda}}
dx-\int_0^{y}\frac{\tilde F(x)}{x^{\lambda}}dx.\Big]
\label{s_y2}
\end{equation}
Here we use the fact that ${\tilde F(x)}\sim x^{m_0}$ as $x\to 0$
and hence
\begin{equation}
{\cal I}(\lambda)\equiv \int_0^{\infty}\frac{\tilde F(x)}{x^{\lambda}} dx
\end{equation}
converges. The last term can now be integrated term by term using the
expression of $\tilde F$. The result is
\begin{equation}
S(y)=(\lambda-1){\cal I}(\lambda)y^{\lambda-1}-
(\lambda-1)\sum_{n=0}^{\infty}\frac{f_n}{n+1-\lambda}y^n.
\label{s_y3}
\end{equation}
Note that $\tilde F$ depends on $m_0$, the integer part of $\lambda$.
When $\lambda=m_0+1$ (integer), we set $\lambda=m_0+1-\epsilon$
in the above formula and let $\epsilon \to 0^+$. In this way,
the singular term obtains a logarithmic factor. The result is
\begin{equation}
S(y)=m_0 {\tilde {\cal I}}y^{m_0}-m_0 f_{m_0} y^{m_0} \ln y
-m_0 \sum_{n=0(\ne m_0)}^{\infty}\frac{f_n}{n-m_0} y^n,
\label{s_y4}
\end{equation}
where
\be
{\tilde {\cal I}}=\int_1^{\infty}\frac{\tilde F(x)}{x^{m_0+1}}dx
+\int_0^1 \frac{\tilde F(x)-f_{m_0}x^{m_0}}{x^{m_0+1}}dx.
\ee
A special case $F(x)=1-\exp(-x)$ has been treated in~\cite{Lee04}.

\section{The leading order analysis of
$\mathcal{B}_{n_{1},n_{2},n_{3},n_{4}}$}

$\mathcal{B}_{n_{1},n_{2},n_{3},n_{4}}$ is defined in
Eq.~(\ref{b_def}) with $M'=(\lambda-2)M/(\lambda-1)$ and
$Q'=(\lambda-2)Q/(\lambda-1)$ and so on.
To see how the leading order behavior of ${\cal B}_{n_1,n_2,n_3,n_4}$
is determined, consider for simplicity the integral
\begin{eqnarray}
{\cal B}_{n_1,n_2,0,0} = (\lambda-1) \int_{1}^{\infty} dz ~
z^{-\lambda} \tanh^{n_{1}} zM' \tanh^{n_{2}} zQ'
\label{c_1}
\end{eqnarray}
with the condition $1 \gg M' \gg Q'$.

When $\lambda$ is sufficiently large, the leading orders in $M'$
and $Q'$ are given by the first terms of the expansion of $\tanh x
= x + \cdots$ and we have
\begin{eqnarray}
{\cal B}_{n_1,n_2,0,0} &\simeq & (\lambda-1) M'^{n_{1}} Q'^{n_{2}}
\int_{1}^{\infty} dz ~ z^{n_{1}+n_{2}-\lambda} \nonumber\\
&=& a_{n_{1}+n_{2}} M^{n_{1}} Q^{n_{2}}.
\label{c_2}
\end{eqnarray}
Eq.(\ref{c_2}) with $a_{l}$ given in Eq.(\ref{a_l}) holds as long as
$\lambda > n_{1} + n_{2} + 1$, but the integral in Eq.(\ref{c_2})
diverges when $\lambda < n_{1} + n_{2} + 1$ indicating appearance
of the non-analytic term as the leading term.

When $\lambda=n_1+n_2+1$, the next leading order in Eq.(\ref{c_2})
cancels the divergence in $a_{n_1+n_2}$ to give
\begin{equation}
{\cal B}_{n_1,n_2,0,0}\approx
(\lambda-1)\Big(\frac{\lambda-2}{\lambda-1}\Big)^{(\lambda-1)}
M^{n_1}Q^{n_2}\ln (1/M).
\label{c_3}
\end{equation}

When $n_{2} + 1 < \lambda < n_{1} + n_{2} + 1$, one scales $z \to
z/M'$ in Eq.(\ref{c_1}) to find
\begin{eqnarray}
{\cal B}_{n_1,n_2,0,0} &=& (\lambda-1) M'^{\lambda-1} \Big\{
\int_{0}^{\infty} dz ~ z^{-\lambda} \tanh^{n_{1}} z \tanh^{n_{2}}
(zQ/M) \nonumber\\
&&~~~~~~~~~~~~~~~~~~~~~~~~
 - \int_{0}^{M'} dz ~ z^{-\lambda} \tanh^{n_{1}} z \tanh^{n_{2}}
(zQ/M) \Big\}.
\label{c_4}
\end{eqnarray}
The second term is $\mathcal{O}(M^{n_{1}+n_{2}+1-\lambda})$ smaller
than the first whose leading contribution is
\begin{eqnarray}
{\cal B}_{n_1,n_2,0,0} &\simeq& (\lambda-1) M'^{\lambda-1}
(Q/M)^{n_{2}}\int_{0}^{\infty} dz z^{n_{2}-\lambda}
\tanh^{n_{1}}z \nonumber\\
&=& \mathcal{C}_{n_{1},n_{2}} M^{\lambda-1-n_{2}}Q^{n_{2}},
\label{c_5}
\end{eqnarray}
where $\mathcal{C}_{n,p}$, defined as
\begin{equation}
\mathcal{C}_{n,p} \equiv
\frac{(\lambda-2)^{\lambda-1}}{(\lambda-1)^{\lambda-2}}
\int_{0}^{\infty} dx~ x^{-\lambda +p} \tanh^{n}x,
\label{c_6}
\end{equation}
converges for $p+1 < \lambda < n+p+1$.

When $\lambda=n_2+1$, similarly to Eq.(\ref{c_3}),
\begin{equation}
{\cal B}_{n_1,n_2,0,0}\approx (\lambda-1)
\Big(\frac{\lambda-2}{\lambda-1}\Big)^{(\lambda-1)}Q^{n_2} \ln (M/Q).
\label{c_7}
\end{equation}
When $1 < \lambda < n_{2}+1$, one scales $z \to z/Q'$ in Eq.(\ref{c_1})
to write it as
\begin{eqnarray}
{\cal B}_{n_1,n_2,0,0} &=& (\lambda-1) Q'^{\lambda-1}
\int_{Q}^{\infty} dz ~ z^{-\lambda} \tanh^{n_{2}} z \tanh^{n_{1}}
(zQ/M).
\label{c_8}
\end{eqnarray}
Since $Q \ll M \ll 1$, $\tanh (zM/Q) \approx 1$ for all $z$ except
near the origin where the contribution to the integral is
negligible. Thus we have
\begin{eqnarray}
{\cal B}_{n_1,n_2,0,0} \simeq (\lambda-1) Q'^{\lambda-1}
\int_{0}^{\infty} dz ~ z^{-\lambda} \tanh^{n_{2}} z =
\mathcal{C}_{n_{2},0} Q^{\lambda-1}.
\label{c_9}
\end{eqnarray}

The leading order terms of $\mathcal{B}_{n_1,n_2,n_3,n_4}$ for
various $\lambda$'s are listed in TABLE III. For simplicity,
we do not show the $\lambda=$ integer cases in TABLE III.
For the border line cases of $\lambda$ dividing the regions of
$\lambda$ with different expressions, a logarithm correction
appears as given in Eq.~(\ref{s_y4}) or (\ref{c_3}) or (\ref{c_7}),
while for other integer values of $\lambda$, the expressions are continuous.

\begin{table}[b]
\begin{tabular}{c|ccccc}
\hline
~~integrals~~ & ~~~~~$2<\lambda<3$~~~~~ & ~~~~~$3<\lambda<4$~~~~~ & ~~~~~$4<\lambda<5$~~~~~ & ~~~~~$5<\lambda<6$~~~~~ & ~~~~~$\lambda>6$~~~~~ \\
\hline
$\mathcal{B}_{2,1,0,0}$ & $\mathcal{C}_{2,1} M^{\lambda-2}Q$ & $\mathcal{C}_{2,1} M^{\lambda-2}Q$ &  $a_{3}M^{2}Q$ & $a_{3}M^{2}Q$ & $a_{3}M^{2}Q$ \\
$\mathcal{B}_{3,0,1,0}$ & $\mathcal{C}_{3,1} M^{\lambda-2}Q_{3}$ & $\mathcal{C}_{3,1} M^{\lambda-2}Q_{3}$ & $\mathcal{C}_{3,1} M^{\lambda-2}Q_{3}$ & $a_{4}M^{3}Q_{3}$ & $a_{4}M^{3}Q_{3}$ \\
$\mathcal{B}_{2,2,0,0}$ & $\mathcal{C}_{2,0} Q^{\lambda-1}$ & $\mathcal{C}_{2,2} M^{\lambda-3}Q^{2}$ & $\mathcal{C}_{2,2} M^{\lambda-3}Q^{2}$ & $a_{4}M^{2}Q^{2}$ & $a_{4}M^{2}Q^{2}$ \\
$\mathcal{B}_{0,3,0,0}$ & $\mathcal{C}_{3,0} Q^{\lambda-1}$ & $\mathcal{C}_{3,0} Q^{\lambda-1}$ & $a_{3}Q^{3}$ & $a_{3}Q^{3}$ & $a_{3}Q^{3}$ \\
$\mathcal{B}_{4,0,0,1}$ & $\mathcal{C}_{4,1} M^{\lambda-2}Q_{4}$ & $\mathcal{C}_{4,1} M^{\lambda-2}Q_{4}$ & $\mathcal{C}_{4,1} M^{\lambda-2}Q_{4}$ & $\mathcal{C}_{4,1} M^{\lambda-2}Q_{4}$ & $a_{5}M^{4}Q_{4}$ \\
$\mathcal{B}_{0,2,0,1}$ & $\mathcal{C}_{2,1} Q^{\lambda-2}Q_{4}$ & $\mathcal{C}_{2,1} Q^{\lambda-2}Q_{4}$ & $a_{3}Q^{2}Q_{4}$ & $a_{3}Q^{2}Q_{4}$ & $a_{3}Q^{2}Q_{4}$ \\
$\mathcal{B}_{0,4,0,0}$ & $\mathcal{C}_{4,0} Q^{\lambda-1}$ & $\mathcal{C}_{4,0;} Q^{\lambda-1}$ & $\mathcal{C}_{4,0} Q^{\lambda-1}$ & $a_{4}Q^{4}$ & $a_{4}Q^{4}$ \\
\hline
\end{tabular}
\caption{The leading order terms of Eq.(\ref{b_def}). Here
$\mathcal{C}_{n,p} \equiv
((\lambda-2)^{\lambda-1}/(\lambda-1)^{\lambda-2})
\int_{0}^{\infty}dx~ x^{-\lambda +p}\tanh^{n}x$, which converges
in the region of $p+1 < \lambda < n+p+1$ and
$a_l=(\lambda-2)^{l}/[(\lambda-1)^{l-1}(\lambda-1-l)]$.}
\end{table}

\end{document}